\documentclass[aps,prb,twocolumn,superscriptaddress,10pt,floatfix]{revtex4-2}
\usepackage[colorlinks=true,citecolor=blue,linkcolor=blue,urlcolor=blue,]{hyperref}
\usepackage{graphicx, nicefrac, textcomp}
\usepackage{float}
\usepackage[normalem]{ulem}
\usepackage{amssymb}
\usepackage{soul}
\usepackage{color}
\usepackage{textcomp}
\DeclareRobustCommand{\textendash}{\ifmmode\text{--}\else\leavevmode\hbox{--}\fi}
\usepackage{graphicx}
\usepackage{gensymb}
\usepackage{graphics}\usepackage{subfigure}\usepackage{pstricks}\usepackage{dcolumn}\usepackage{bm}
\usepackage{siunitx}
\usepackage{booktabs}
\usepackage{multirow}
\usepackage{siunitx}

\begin{document}

\title{Topochemical Fluorination of La$_2$NiO$_{4+\delta}$ Single Crystals}

\author{H.~Yilmaz}
\affiliation{University of Stuttgart, Institute for Materials Science, Materials Synthesis Group, Heisenbergstra{\ss}e 3, 70569, Stuttgart, Germany}
\affiliation{Turkish Energy, Nuclear and Mineral Research Agency, Nuclear Energy Research Institute, Kahramankazan, Ankara, 06980, Türkiye}
\author{M.~Isobe}
\affiliation{Max Planck Institute for Solid State Research, Heisenbergstra{\ss}e 1, 70569, Stuttgart, Germany}

\author{O.~Clemens}
\email{oliver.clemens@imw.uni-stuttgart.de}
\affiliation{University of Stuttgart, Institute for Materials Science, Materials Synthesis Group, Heisenbergstra{\ss}e 3, 70569, Stuttgart, Germany}

\author{P.~Puphal}
\email{puphal@fkf.mpg.de}
\affiliation{Max Planck Institute for Solid State Research, Heisenbergstra{\ss}e 1, 70569, Stuttgart, Germany}

\date{\today}

\begin{abstract}

Topochemical fluorination offers a low-temperature route for modifying the anion chemistry and electronic ground states of layered transition-metal oxides, providing access to metastable phases and functionalities that are not able to be achieved through conventional solid-state synthesis. Despite extensive work on polycrystalline samples and thin films, topochemical fluorination of bulk single crystals has not been studied, limiting insights into intrinsic structure–property relationships. Here, we investigate the topochemical fluorination of optical float zone–grown (OFZ) La$_2$NiO$_{4+\delta}$ single crystals using polymer-based PTFE-Polytetrafluoroethylene $(\mathrm{C}_{2}\mathrm{F}_{4})_{n}$, PVDF-Polyvinylidene fluoride $(\mathrm{C}_{2}\mathrm{H}_{2}\mathrm{F}_{2})_{n}$ and inorganic ($\mathrm{CuF}_{2}$) fluorination agents and compare it to our topochemical pathways of reduction of LaNiO$_{3-x}$. By systematically investigating direct and indirect contact reaction pathways, we can understand fluorination mechanisms, quantify the degree of fluorine incorporation, and evaluate the resulting structural and magnetic modifications in a detail that was not possible in powder and thin films. Powder and single-crystal X-ray diffraction reveal that fluorination proceeds without destroying the Ruddlesden–Popper framework, while inducing lattice parameter changes consistent with anion intercalation in the bulk and ion exchange on the surface. This even induces a clear superstructure, which was not reported before and extends the understanding of anion insertion reactions beyond what is known on stage ordering in nickelates. Energy-dispersive X-ray spectroscopy confirms strong fluorine incorporation on the surface and reduced homogeneity in the bulk. Magnetic susceptibility measurements demonstrate a change in antiferromagnetic ordering upon fluorination. These findings highlight the potential of topochemical fluorination as a flexible post-growth approach for tailoring magnetic and electronic properties even for single crystals of layered nickelates, with applications in functional oxide design for post growth doping to tune a novel class of superconductors.

\end{abstract} 

\maketitle

\section{Introduction}
Layered Ruddlesden–Popper (RP)-type oxides (general formula $A_{n+1}B_{n}O_{3n+1}$) are essential materials to investigate the interaction between lattice, charge and spin degrees of freedom, owing to their anisotropic crystal chemistry and the presence of accessible interstitial sites within the rock-salt layers \cite{Yang2025_ncomm_3277,Zhang2016,Herlihy2025,Flathmann2024_APLMat_061112,Balachandran2014}. In these structures, the substitution or intercalation of anions can significantly influence the electronic bandwidth and super-exchange mechanisms, thereby adjusting the magnetic and transport properties without disrupting the primary perovskite framework. Anion topochemistry, particularly the insertion of fluorine into oxide lattices, has emerged as a promising method of modifying the lattice and valence states of transition metals at significantly lower temperatures than those required for conventional high-temperature synthesis \cite{Vanita2024,Vanita2025,Yilmaz2025_CommPhys_408,Wissel2020,wissel_LNOF,Puphal_LNO2,Puphal2025_NatRevPhys}. For example, topochemical fluorination of the bilayer iridate Sr$_3$Ir$_2$O$_7$ single crystals transforms it into Sr$_3$Ir$_2$O$_7$F$_2$, in which the oxidation state of iridium (Ir) changes from 4+ to 5+, thereby converting the material from a relativistic Mott insulator to a spin–orbit-driven band insulator \cite{Peterson_Sr3Ir2O7F2}. In another example, the insertion of F ions into NdNiO$_{2+x}$ produced a mixed anion perovskite with unique octahedral tilting patterns \cite{Yang2025_ncomm_3277}. It has also been shown that the fluorination of RP oxides can produce new magnetic or electronic ground states. This phenomenon emphasises how F$^{-}$ insertion fundamentally modifies correlated electronic states and magnetic order. In bilayer RP oxides, Sr$_3$(M$_{0.5}$Ru$_{0.5}$)$_2$O$_7$ (M~=~Ti, Mn, Fe) reacts with CuF$_2$ to produce oxyfluorides Sr$_3$(M$_{0.5}$Ru$_{0.5}$)$_2$O$_7$F$_2$, where Ru is oxidised to Ru$^{6+}$. These F-O exchanges have a significant impact on magnetism: the fluorination of Sr$_3$(Ti$_{0.5}$Ru$_{0.5}$)$_2$O$_7$ to Sr$_3$(Ti$_{0.5}$Ru$_{0.5}$)$_2$O$_7$F$_2$ suppresses magnetic order as the compound has increased conductivity \cite{Fabio_2013}. Overall, the process of mixed anion chemistry allows for exceptional tunability and structural control \cite{Kageyama2018}.\\
La$_2$NiO$_{4+\delta}$ phase, which belongs to the Ruddlesden–Popper n=1 nickelate series, is a well-known layered antiferromagnet in which the electronic and magnetic properties are closely associated with the anion content. The structure consists of NiO$_{4/2}$O$_2$ planes, in between which La$^{3+}$ ions are located. rock-salt layers that host excess oxygen in interstitial sites, resulting in mixed Ni$^{2+}$/Ni$^{3+}$ valence states. Variations in oxygen excess are known to considerably affect the Néel temperature, magnetic anisotropy and in-plane electrical conductivity by modifying the carrier concentration and superexchange interactions within the NiO$_2$ planes \cite{Acrivos1994,Gopalan.45.249,PAULUS2002565}. Consequently, La$_2$NiO$_{4+\delta}$ has been widely used as a model system to study the interaction between lattice distortions, charge doping and magnetism in layered transition-metal oxides.\\
As discussed above, an alternative route to tailor anion composition is provided by polymer-assisted fluorination of La$_2$NiO$_{4+\delta}$, which results in the formation of the oxyfluoride La$_2$NiO$_3$F$_2$, with an unconventional oxide–fluoride anion ordering and a pronounced suppression of the antiferromagnetic ordering temperature (650~K in La$_2$NiO$_4$ \cite{Lander1989} to about 49~K in La$_2$NiO$_3$F$_2$) \cite{wissel_LNOF}. These studies demonstrate that controlled fluorination via topochemical pathways can be used to modify the structure and physical properties of layered oxide systems.
Additional topochemical reduction was shown to successfully reduce the system to La$_2$NiO$_3$F \cite{Wissel2020_ChemMat_3160}, for which theoretical investigation indicate promising analogy of cuprate-like physics while keeping distinct Ni$^{1+}$
features \cite{Bernardini2021,Harada2024}. Notably in this regard doped infinite-layer Ni$^{1+}$ was shown to superconduct \cite{Li2019} and since this discovery various nickelate superconductors were shown to exist \cite{Puphal2025_NatRevPhys}. 
\\
Up to now, many existing studies have focused on polycrystalline samples or thin films in context of topochemical fluorination. This means that the intrinsic structure–property relationships can be difficult to understand due to the presence of grain boundaries, strain and compositional inhomogeneity. In contrast, bulk single crystals provide a valuable opportunity to investigate intrinsic anisotropic properties, enabling direct correlations to be established between crystallographic modifications and macroscopic physical behaviour. In addition single crystals enable additional characterization techniques, thereby allowing impurity-related properties to be resolved due to the possibility of spatial separation and advanced microscopy. Despite these advantages, there have been no systematic studies of topochemical fluorination in La$_2$NiO$_{4+\delta}$ single crystals, and fundamental questions regarding fluorination pathways, achievable fluorine content and their influence on magnetism and electrical conductivity need to be answered.\\
The ability to fine-tune the electronic and magnetic properties of quantum materials is central to the discovery of novel quantum phenomena. In nickelates, such as LaNiO$_3$, post-growth chemical modifications offer a promising route to access new phases without altering the bulk crystal structure. Recent advances in the topochemical reduction of La$_2$NiO$_{4+\delta}$ have demonstrated the feasibility of this approach on both microscale flux-grown crystals \cite{Puphal2021} and millimeter-sized single crystals grown by the optical floating zone method \cite{Puphal_LNO2}. These studies have revealed a rich reduction pathway involving the formation of novel structural phases \cite{Wu2023}, as well as the emergence of metallic behavior, distinct from the insulating character typically observed in polycrystalline samples \cite{Puphal2021,Puphal_LNO2,Suyolcu2025}. Most importantly with optimal doping a superconducting state is created \cite{Li2019}. Furthermore, high-resolution microscopy on reduced single crystal has enabled the direct observation of unconventional interfacial structures formed during the reduction process \cite{Wu2024}.

However, a key challenge remains: the removal of one apical oxygen atom per formula unit can lead to the formation of split domains, even resulting in physical separation of crystal regions \cite{Puphal2021}, which compromises structural and electronic homogeneity. This limitation underscores the need for alternative, more controlled chemical transformations.

Motivated by these considerations, we herein investigate the topochemical fluorination of optical floating zone-grown La$_2$NiO$_{4+\delta}$ single crystals using a variety of fluorination agents. Our primary goals are to understand the fluorination mechanisms, quantify the degree of fluorine incorporation, and systematically examine the resulting changes in magnetic properties. Through the combination of controlled topochemical reactions and comprehensive structural and magnetic characterization, we establish fluorination as a viable and tunable post-growth approach for adjusting the physical properties of layered nickelate single crystals. This work not only expands the toolkit for defect engineering in quantum materials but also emphasizes the critical role of synthesis control in achieving reproducible and well-defined functional states.

\begin{figure*}[hbt]
\centering
\includegraphics[width=2\columnwidth]{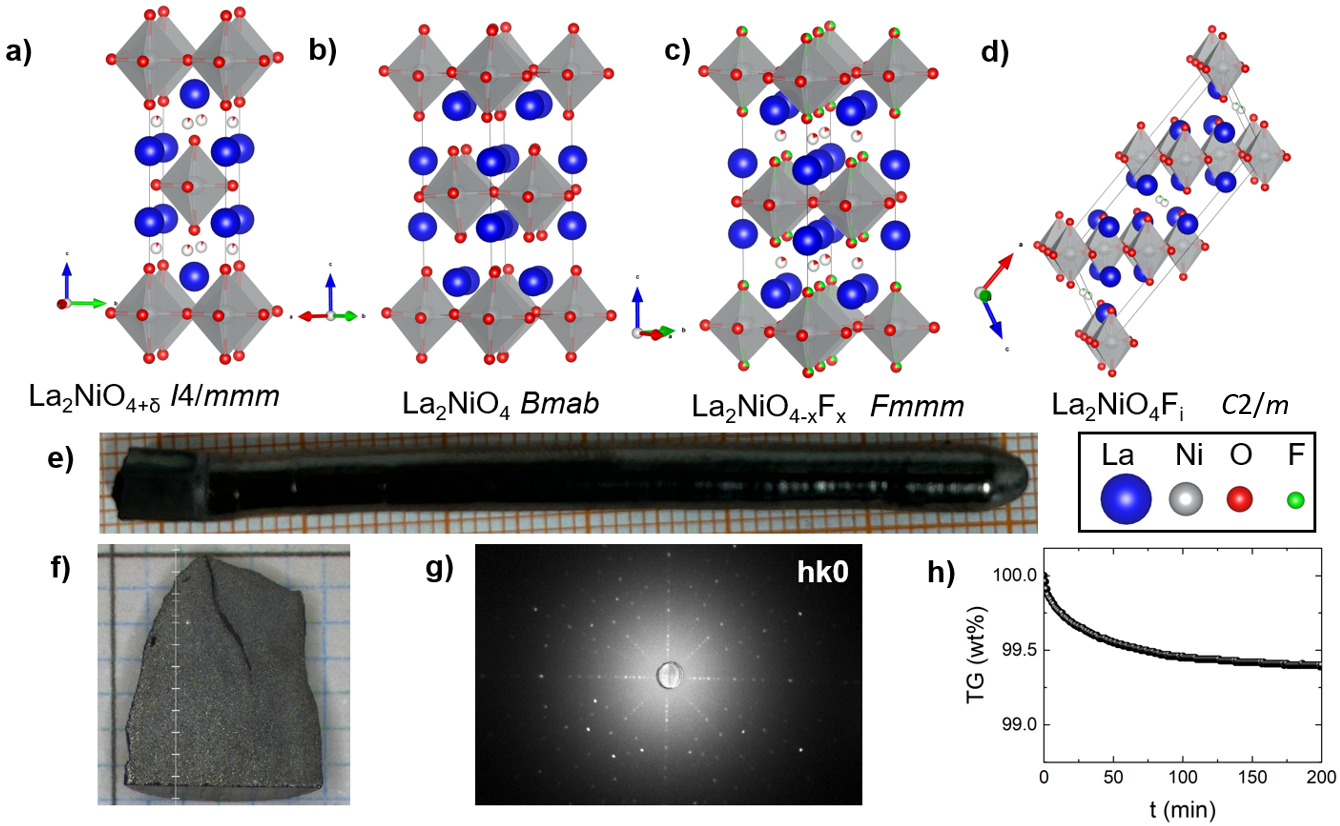}
\caption{\textbf{Structure.} Crystal structure, diffraction, and macroscopic appearance of the grown oxide single crystal. (a) Crystal structure of La$_2$NiO$_{4+\delta}$ (\textit{I}4/\textit{mmm}) (for $0.1<\delta<0.15$  \cite{Tamura1993} (b) reduced La$_2$NiO$_{4}$ (\textit{Bmab}) (c) La$_2$NiO$_{4-x}$F$_{2x}$ (\textit{Fmmm}) and (d) a Fluorine intercalated superstructure discussed further below of La$_2$NiO$_{4}$F$_{i}$ (\textit{C}2/$m$), all highlighting the corner-sharing octahedral framework. (b) Laue X-ray diffraction pattern recorded in the $hk0$ plane, confirming high crystalline quality and long-range order. (e) As-grown rod and (f) cut single crystal pieces obtained by the optical floating zone method. (g) Laue X-ray diffraction pattern recorded in the $hk0$ plane, confirming high crystalline quality and long-range order. (h) TGA reduction measurement in a flow of forming gas 95\%Ar/5\%H$_2$ of 100~cc at 450\degree C over time.}
\label{Crystal_structure}
\end{figure*}

\section{Results}
\subsection{Structural Characterization by Powder and Single-Crystal XRD}
The crystal structure, Laue diffraction, and macroscopic appearance of the as grown oxide single crystal are displayed in Fig.~\ref{Crystal_structure}. Fig.~\ref{Crystal_structure}~(a) shows the crystal structures of the parent Ruddlesden–Popper phase La$_2$NiO$_{4+\delta}$, crystallizing in the tetragonal \textit{I}4/\textit{mmm} space group with randomly distributed interstitial oxygen at a low level without superstructure ordering. The topochemically reduced phase La$_2$NiO$_4$, which adopts an orthorhombic \textit{Bmab} symmetry showing octahedral tilting and is shown in Fig.~\ref{Crystal_structure}~(b). 

 Fig.~\ref{Crystal_structure}~(e) shows the as-grown crystal rod obtained by the optical floating zone method (see methods section), along with a representative cut single-crystal piece (f). The uniform appearance of the growth rod and the ability to obtain large, well-faceted fragments further attest to the suitability of the crystals for detailed structural characterization and subsequent physical property measurements. 
 Fig.~\ref{Crystal_structure}~(g) presents a Laue X-ray diffraction pattern recorded in the $hk0$ plane from a representative single-crystal specimen. The presence of sharp, well-defined, and symmetrically arranged diffraction spots, together with the absence of streaking or splitting, demonstrates the high crystalline quality of the crystal and confirms the establishment of long-range structural order across the probed volume.
 Due to the synthesis condition of the rods excess oxygen is in the obtained boule even with argon atmosphere during growth. We determine the oxygen content via thermogravimetric analysis (TGA); as shown in Fig.~\ref{Crystal_structure}~(h) a mass loss of 0.61 wt\% is observed leading to an excess estimation of La$_2$NiO$_{4.15}$.

\begin{figure}[t]
\centering
\includegraphics[width=1.0\columnwidth]{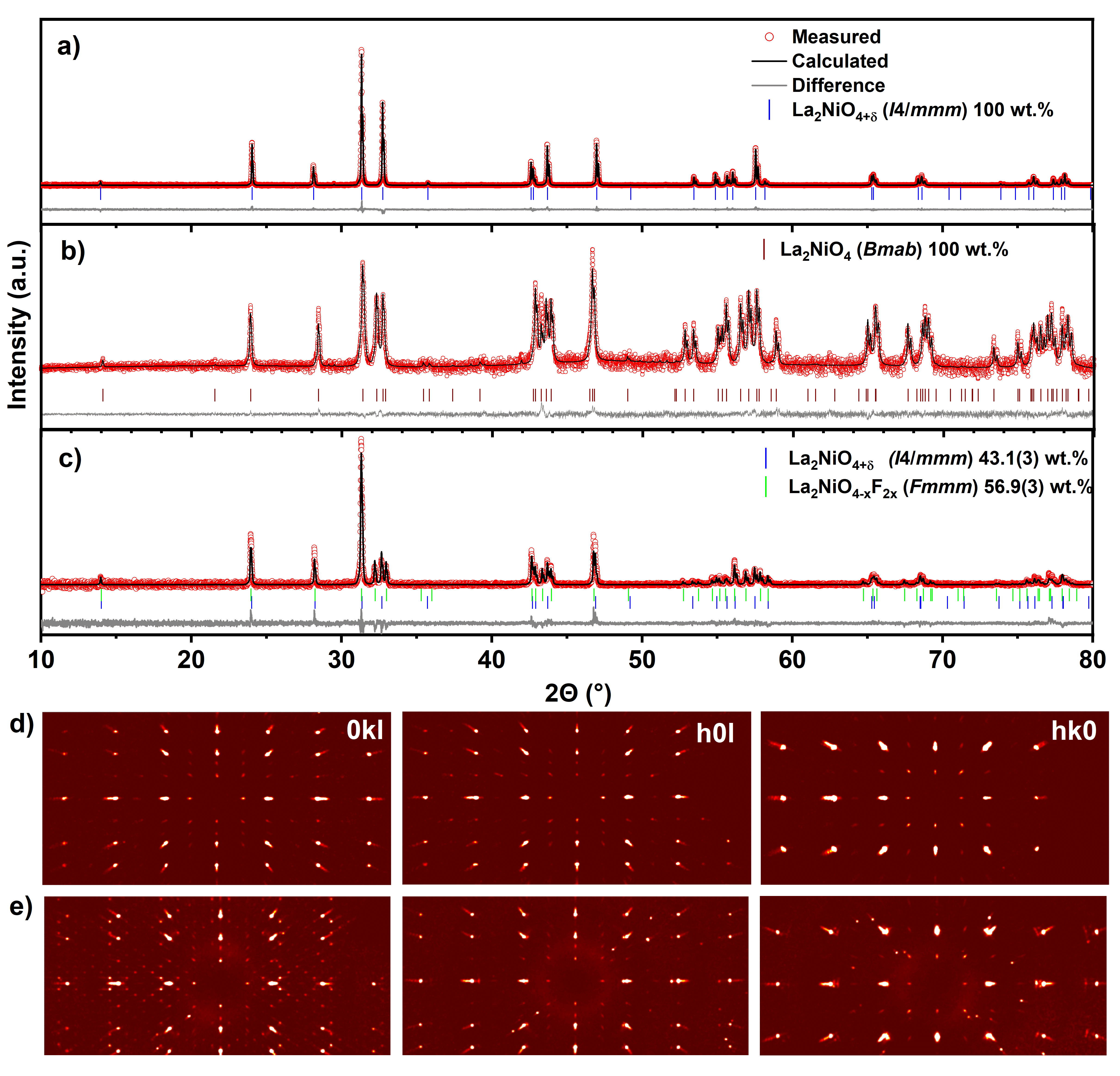}
\caption{\textbf{XRD.} Rietveld fit of the room-temperature powder X-ray Diffraction (\textbf{PXRD}) pattern of (a) as-grown La$_2$NiO$_{4+\delta}$ (b) after TGA reduction in Ar/H$_2$ flow at 450~$\degree$C (c) after first PTFE fluorination. Rietveld fit of the room-temperature Single Crystal X-ray Diffraction (\textbf{SC-XRD}) pattern of (d) as-grown La$_2$NiO$_{4+\delta}$, (e) after twice PTFE fluorination treatment. 
}
\label{PowderXRD}
\end{figure}

The crystal structure of as grown powdered La$_2$NiO$_{4.15}$ single crystals have been determined by Rietveld refinement of the X-ray diffraction pattern, as shown in Fig.~\ref{PowderXRD}~(a). As clearly seen from the refinement, phase pure formation of La$_2$NiO$_{4+\delta}$ has been achieved. La$_2$NiO$_{4+\delta}$ crystallizes in a perovskite-related Ruddlesden--Popper structure, belonging to the tetragonal crystal system, (space group \textit{I}4/\textit{mmm}), with lattice parameters a=b= 3.864874(12)~\AA, c = 12.67435(5)~\AA~and V = 189.3199(14)~\AA$^{3}$. These structural parameters are in good agreement with previously reported literature values \cite{Nowroozi2020_CommMat_27,RodriguezCarvajal1991_JPCM_3215}.\\

After topochemical reduction, e.g. within the thermogravimetry measurements (shown in Fig. \ref{Crystal_structure} (h), the crystal adopts the  La$_2$CuO$_4$ type structure, (space group \textit{Bmab}), with lattice parameters a = 5.536708(9)~\AA, b = 5.467154(11)~\AA, c = 12.54052(5)~\AA, and V = 379.6021(14)~\AA$^{3}$ (refinement in Fig.~\ref{PowderXRD}~(b)).
After topochemical fluorination of La$_2$NiO$_{4+\delta}$ single crystal, a significant structural change was found as indicated on a once PTFE treated crushed crystal shown in X-ray diffraction (PXRD) and corresponding Rietveld refinement (see Fig.~\ref{PowderXRD}~(c)). Our XRD pattern revealed significant peak splitting and intensity changes, indicating the formation of a two-phase material, possibly related to limited reaction depth on the single crystal. Rietveld refinement confirmed the coexistence of 43.1(3)~wt.\% of a tetragonal \textit{I}4/\textit{mmm} phase with slightly expanded a/b parameters and contracted c axis (a~=~b =~3.87054(6)~\AA, c~=~12.6315(3)~\AA), together with 56.9(3)~wt.\% of an orthorhombic \textit{Fmmm} phase characterized by lattice parameters a~=~5.54769(10)~\AA, b~=~5.42446(10)~\AA, and c~=~12.6326(3)~\AA. The emergence of the orthorhombic \textit{Fmmm} phase (56.9(3)~wt.\%) reflects a lowering of structural symmetry, which is commonly associated with anion ordering phenomena, partial anion substitution, or fluorine intercalation. Such structural modifications are likely to occur within the rock-salt layers of Ruddlesden–Popper-type compounds during topochemical fluorination.
Notably, various phases upon fluorination have been discussed in powder form \cite{Jacobs2025} with a extreme fluorination up to  La$_2$NiO$_{2.5}$F$_3$.
The established structure extracted from powder XRD of ion exchange with fluorine substituting the apical side and pushing oxygen into the interstitials in the orthorhombic \textit{Fmmm} structure is shown in Fig.~\ref{Crystal_structure}~(c) \cite{wissel_LNOF,Jacobs2024}.  All structures are based on a framework of corner-sharing NiO$_6$ octahedra; however, the topochemical fluorination leads to a symmetry lowering and pronounced modifications of the local coordination environment, reflecting the ordered incorporation of fluorine ions and the associated rearrangement of the anion sublattice. For full fluorination a clear indication of the intercalated cations is the change of the surrounding octahedrons leading to the tilting of the octahedra between neighboring layers switching from in phase (La$_2$NiO$_4$) to antiphase (La$_2$NiO$_3$F$_2$) \cite{wissel_LNOF}. This leads to a change of a seemingly kite square environment around the interstitials to an elongated hexagon (see Fig. 9 in \cite{wissel_LNOF}).  
Besides this fluorine ion exchange, fluorine intercalation was also studied on powder samples via electrochemical charging \cite{Nowroozi2020_CommMat_27}. Here, starting from La$_2$NiO$_{4.13}$ fluorine fills the interstitial tetrahedral fluorite slabs in the rocksalt LaO layers up to  La$_{2}$NiO$_{4.13}$F$_{1.87}$. In this case, no considerable tiltings are observed but a continuous change of the $a,b$ lattice constant difference and $c-$axis increase is observed ranging from 12.671(1) to 12.761(4) \AA, for fluorine free and full intercalation respectively.

Our Powder XRD results indicate that fluorination has occurred to a certain degree throughout the crystal volume in the first cycle, but these results are not sufficient for a detailed understanding of the fluorination mechanism.\\

Besides PXRD measurements, Single-crystal XRD measurements were performed at ambient temperature before and after fluorination to additionally confirm the targeted composition and structure. Fig.~\ref{PowderXRD}~(d), and (e) presents the single-crystal diffraction patterns of as grown La$_2$NiO$_{4+\delta}$ and twice PTFE-fluorinated La$_2$NiO$_{4-x}$F$_{2x}$ both integrated with the same \textit{Fmmm} setting for comparison. Consistent with Fig.~\ref{PowderXRD}~(a), the parent La$_2$NiO$_{4+\delta}$ phase displays well-defined diffraction patterns with sharp, symmetric reflections, particularly in the $hk0$ plane. These features are characteristic of a high-symmetry tetragonal structure consistent with the space group \textit{I}4/\textit{mmm} (\# 139), with refined lattice parameters a~=~b~=~3.8977(6)~\AA, c~=~12.535(3)~\AA, and a unit-cell volume of V~=~190.43(6)~\AA$^{3}$. No evidence of superlattice reflections or peak splitting was observed, confirming the absence of long-range octahedral tilting or anion sublattice ordering. This observation is characteristic of an ideal Ruddlesden–Popper n~=~1 structure, in which corner-sharing NiO$_6$ octahedra extend within the ab plane and successive perovskite layers are separated by intervening LaO rock-salt layers and intercalated oxygen atoms are too disordered to create superstructure patterns.\\

After the second fluorination step, the Single-Crystal XRD patterns exhibit well-defined superstructure reflections, without strong diffuse scattering or peak multiplicity that was observed at an intermediate stage (not shown). These superstructure reflections break the symmetry along two directions and only maintain a two-fold symmetry along the [110]tet direction. Disregarding these superstructure reflections, the diffraction data can be fully indexed within an orthorhombic \textit{Fmmm} structure.
Considering the basic \textit{Fmmm} cell the lattice parameters (not accounting for the superstructure reflexes) are $a$~=~5.4135(5)~\AA, $b$~=~5.5539(5)~\AA, $c$ = 12.6574(12)~\AA, with a unit-cell volume of V = 379.99(19)~\AA$^{3}$.  If we use the lattice constants change in relation to the amount of fluorine incorporation characterized in Ref. \cite{Jacobs2025} the difference in a and b is closest to the inter 2 phase with \textit{Fmmm}. Similarly when comparing to the $c-$axis change established in Ref. \cite{Nowroozi2020_CommMat_27} this corresponds to a low fluorine incoroporation level at the range of 10\%. 

If we now consider the additional superstructure reflexes in a refinement, we find a tripling of the $b-$lattice parameter with 16.6661(14)~\AA, while a=5.4136(4)~\AA stays. Notably, the map shown in Fig. \ref{PowderXRD} (e) also has an l component with intermediate reflexes along the height at around half of the main bragg reflexes. For the refinement we have to consider that  F has only a ~9.7\% higher scattering factor than O and its relative refinement requires high resolution data, where single crystal diffraction allows to refine atomic displacement parameters (Uani) and occupancies to a better resolution. However, our data was acquired with a very simple lab based benchtop XRD that due to limited beam alignement has intrinsic reflex smearing \cite{Puphal2023,Krieger2025}. The structure was solved with a monoclinic cell of $C_2/m$ and the lattice constants $a$~=~16.836(2)~\AA, $b$~=~5.4112(4)~\AA~ and $c$~=~6.9079(8)~\AA, where $\beta=115.019(15)^\circ$. We attempted to refine partial F to all oxygen sites as a substituent, which led to an increase in R value aside from the intercalated site, where it improved the overall refinement when placing F. The refinement results are summarized in Tab. \ref{sc}. The final resulting structure is displayed already in \ref{sc} (d), showing clear chains of intercalated fluorine along the diagonals between the octahedron on every third space. We find that these sites are partially occupied and considering all XRD results we conclude in bulk an intercalation process of only F$_{0.1}$ fluorine at the tetrahedral interstitial sites of the LaO rocksalt layers.

\begin{table}[htbp]
\small 
\centering
\caption{Crystallographic parameters of the refined structure of La$_2$Ni$_4$O$_4$F$_{0.1}$ with $a$~=~16.836(2)~\AA, $b$~=~5.4112(4)~\AA and $c$~=~6.9079(8)~\AA, where $\beta=115.019(15)^\circ$,  with a goodness of refinement of R=0.0410 \& wR= 0.1096.}
\label{tab:crystallographic}
\begin{tabular}{lcccccc}
\toprule
Atom & $x$ & $y$ & $z$ & $U_{\text{ani}}$ (\si{\angstrom^2}) & Occ. \\
\midrule
La1 & 0.62114(5) & 0.500000 & 0.51889(10) & 0.0145(4) & 1.0 \\
La2 & 0.54663(4) & 0.000000 & 0.81713(10) & 0.0143(4) & 1.0 \\
La3 & 0.78741(5) & 0.000000 & -0.14941(12) & 0.0148(4) & 1.0 \\
Ni1 & 0.500000 & 0.500000 & 0.000000 & 0.0125(7) & 1.0 \\
Ni2 & 0.66648(9) & 0.000000 & 0.3339(2) & 0.0134(5) & 1.0 \\
O1  & 0.5594(6) & 0.500000 & -0.2285(15) & 0.041(3) & 1.0 \\
O2  & 0.7243(6) & 0.000000 & 0.1004(14) & 0.028(2) & 1.0 \\
O3  & 0.6092(6) & 0.000000 & 0.5699(16) & 0.045(3) & 1.0 \\
O4  & 0.5821(3) & 0.2514(15) & 0.1688(7) & 0.0171(16) & 1.0 \\
O5  & 0.750000 & 0.250000 & 0.500000 & 0.014(2) & 1.0 \\
F1  & 0.500000 & -0.256(18) & 0.500000 & 0.011(16) & 0.10 \\
\bottomrule
\label{sc}
\end{tabular}
\end{table}

\subsection{Electron microscopy and Elemental Analysis via Energy-Dispersive X-ray Spectroscopy (EDX)}
\begin{figure}[H]
\centering
\includegraphics[width=0.9\columnwidth]{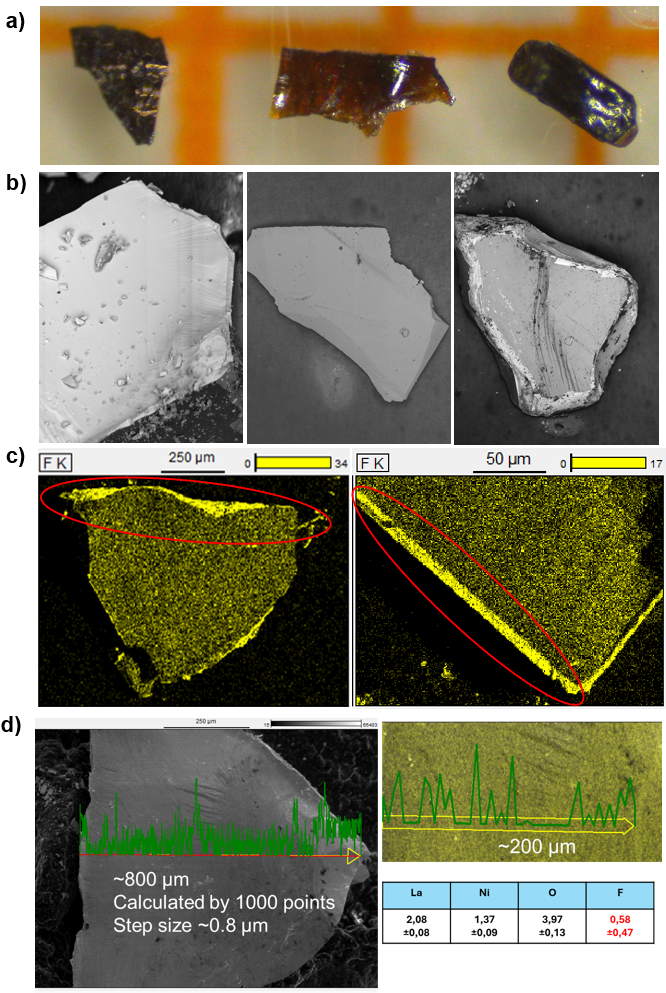}
\caption{\textbf{Chemical composition and distribution.} (a) Microscopic images of La$_2$NiO$_4$ single crystals on millimeter scale before and after TGA reduction in Ar/H$_2$ flow at 450~$\degree$C as well as PTFE treated (last). (b) SEM-BSE images of cross-sectioned single crystals of the same as grown La$_2$NiO$_{4.15}$, reduced La$_2$NiO$_4$ and after PTFE fluorination. (c) Corresponding fluorine (F K$_{\alpha}$) elemental maps on a PTFE treated specimen, highlighting pronounced fluorine accumulation near the crystal surface (marked by red ellipses), indicative of surface-dominated fluorination (d) EDX line scan acquired across the crystal cross section of a cleaved crystal over, as well as a magnified view, of the near-surface region ($\approx$ 200~\si{\micro\metre}), where particularly large variations in fluorine content are observed. with accompanying compositional table
}
\label{SEM_fig}
\end{figure}

The topochemical fluorinated crystals were transferred to a scanning electron microscope (SEM) and its images reveal a way better crystallinity in comparison to the topochemically reduced samples of LaNiO$_{3-x}$ \cite{Puphal2021,Puphal_LNO2,Suyolcu2025}. Unlike the deintercalation, which seperates micrometer sized domains \cite{Hayashida2024,Wu2024} that are visible as cracks and furrows in electron microscopy, the ion exchange has no influence on the crystalline topography. The cut surface look the same as before as visible in Fig.~\ref{SEM_fig}. Even the cleaving of the specimen can be performed more effectively than in the untreated sample. In Figure \ref{SEM_fig} (a) we show optical images of as grown, topochemically reduced and topochemically fluorinated crystals. The reduced crystals become optically transparent with a brown colour, while as grown and fluorinated remain black. Backscattered electron imaging was performed similarly on as grown, reduced and topochemical fluorinated crystals as shown in Fig.\ref{SEM_fig} (b). The elemental mapping allows to see impurities or inhomogeneities. However, all crystals are quite pure, only the surface of fluorinated crystals show some contrast difference.

Energy-dispersive X-ray spectroscopy (EDX) was investigated to estimate the degree of fluorination in the topochemically treated single crystals. The stoichiometry of the as-grown single crystals was determined as La$_{1.98(7)}$Ni$_{0.93(3)}$O$_{4.22(9)}$ by EDX before the topochemical fluorination process. After the fluorination process, the crystals were cleaved perpendicular to the surface, and EDX measurements were conducted across the exposed cross-section. The degree of fluorination was measured on the cross sectional area. To ensure better conductivity, the crystals were embedded in carbon paste and no coating process was carried out on the surface to ensure a more accurate analysis of the elemental sensitivity. Nevertheless, the intrinsic limitations of EDX spectroscopy must be considered. The technique is less sensitive to light elements with low atomic numbers, only provides approximate compositional information. Moreover, its limited spectral resolution can lead to overlap of characteristic X-ray peaks from different elements. Additionally, its limited penetration depth (generally, 1–3~\si{\micro\metre} for heavier elements and 0.5–1~\si{\micro\metre} for lighter elements) limits analysis to the near-surface region of the specimen, which can affect insights into the composition with respect to the bulk crystal \cite{Scimeca2018_EJH_2841}. Nonetheless, we use the technique on comparable single crystals and thus can quite well estimate even oxygen and fluorine contents in our crystals.\\

Comparing fluorination agents: using PVDF as a fluorine source, the fluorine content increased with process temperature, reaching values of 0.85$~\pm~$0.11 per formula unit at 370\degree C and 0.92$~\pm~$0.08 per formula unit at 400\degree C. This trend reflects enhanced fluorine mobility driven by thermally activated diffusion processes. In contrast, fluorination using CuF$_2$ at 250\degree C resulted in a comparatively low fluorine concentration of 0.32$~\pm~$0.08 per formula unit, whereas treatment at 400\degree C yielded a substantially higher incorporation level of 1.02$~\pm~$0.15, comparable to that achieved with PVDF. It should be noted that CuF$_2$-treated samples required ultrasonic cleaning prior to analysis due to surface reactions leading to CuO formation. Fluorine incorporation was found to be most significantly enhanced by fluorination treatments using PTFE (Fig.~\ref{SEM_fig}~(a)). Cross-sectional EDX analysis revealed that a double PTFE treatment increased the fluorine content on the surface from 0.91$~\pm~$0.20 to 2.64$~\pm~$0.18, consistent with a diffusion-limited incorporation mechanism in which repeated treatments facilitate deeper penetration of fluorine into the lattice. In Fig.~\ref{SEM_fig}, EDX elemental maps and line-scan analysis illustrate the distribution of fluorine in La$_2$NiO$_{4+\delta}$ single crystals after topochemical fluorination with PTFE, from the crystal surface toward the bulk on a cleaved surface. This is a minority phase and will be averaged out in single crystal diffraction and major part of it might be amorphous from decomposition. \\
The fluorine intercalation mechanism during topochemical fluorination of single crystals is schematically illustrated by the EDX elemental maps in Fig.~\ref{SEM_fig}~(c). As commonly observed in fluorination processes, the fluorine source decomposes at agent-specific temperatures, releasing F$_2$/CF/HF in the gaseous phase. The generated fluorine initially accumulates at the crystal surface, leading to a surface-dominated reaction regime. This behavior is characteristic of diffusion-controlled process, in which fluorine ions preferentially react at the surface before progressively penetrating into the crystal interior. With continued fluorination, fluorine gradually diffuses more uniformly into the bulk, indicating a transition from surface-limited to bulk diffusion. This evolution of the transport mechanism is likely facilitated by the generation of anion vacancies or interstitial sites within the lattice, which enhance fluorine ion mobility and enable deeper incorporation into the crystal structure besides the intercalated sites between the rocksalt layers visualized in Fig.~\ref{Crystal_structure} creating transport channels. Following this observation from EDX elemental mapping that fluorine is inhomogeneously distributed and accumulates near the crystal surface, EDX line-scan analyses were conducted to further investigate the fluorine diffusion behavior from the surface into the bulk of the crystal. The line-scan data presented in Fig.~\ref{SEM_fig}~(c) provide clear evidence of significant local inhomogeneity in the fluorine distribution within the fluorinated La$_2$NiO$_{4+\delta}$ single crystal. The scan covers $\approx$ 800~\si{\micro\metre} with a high spatial resolution of 0.8~\si{\micro\metre} per step and reveals a strongly non-uniform fluorine profile, where the light elemental character and consequently high systematic error play a crucial role, characterized by large and recurrent fluctuations in atomic percent along the scan direction (Fig.~\ref{SEM_fig}~(c)-right). Nonetheless, the near-surface region extending $\approx$ 200~\si{\micro\metre} from the surface (Fig.~\ref{SEM_fig}~(c)-left) exhibits especially high fluorine concentrations. This behavior suggests that fluorine incorporation during the topochemical fluorination process is not spatially well controlled. Consistent with this observation, the compositional data summarized beneath Fig.~\ref{SEM_fig}~(c) show an exceptionally large deviation in fluorine stoichiometry (0.58$~\pm~$0.47), compared to point measurements on cleaved surfaces with even distance to the surface, providing further evidence that the fluorine content varies substantially throughout the crystal. The challenge of achieving uniform fluorination is revealed by the significant inhomogeneity identified in the elemental distribution and line scan analysis, especially near the crystal surface as well as crystal bulk. This inhomogeneity may affect the characterization of the intrinsic properties of fluorinated crystals and must be taken into account in further structural and physical investigations.

\begin{table}[t]
\centering
\caption{\textbf{Chemical compositions of La$_2$NiO$_{4+\delta}$-based samples prepared by different methods.} 
The compositions are reported with stoichiometric deviations in oxygen ($\mathrm{O}$) and fluorine ($\mathrm{F}$) determined by EDX, revealing the details of topochemistry of reduction, and fluorination treatment. Uncertainties via standard-deviation are given in parentheses and refer to the last digit.}
\label{tab:composition}
\begin{tabular}{lcccc}
\hline
Sample & \multicolumn{1}{c}{La} & \multicolumn{1}{c}{Ni} & \multicolumn{1}{c}{O} & \multicolumn{1}{c}{F}  \\
\hline
Reduced & 2.08(7) & 0.98(4) & 4.04(9) & 0.05(1) \\
As-grown & 1.98(7) & 0.93(3) & 4.22(9) & 0.03(1) \\
Fluorinated, cleave & 2.0(2) & 0.97(8) & 3.9(3) & 0.3(2)  \\
\hline
\end{tabular}
\end{table}

\subsection{Magnetic Characterization of Fluorinated La$_2$NiO$_4$}

\begin{figure}[t]
    \centering
    \includegraphics[width=1.0\columnwidth]{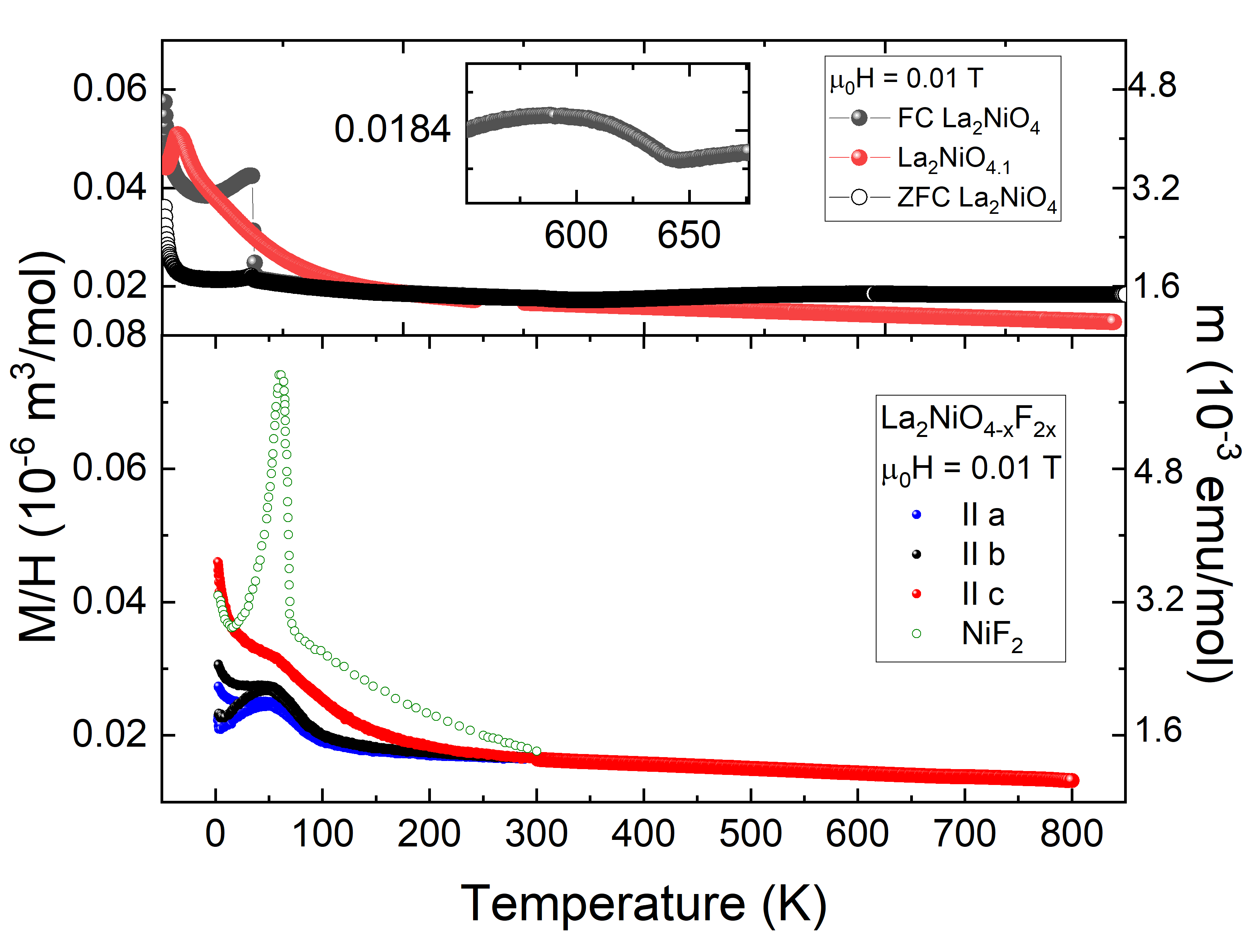}
    \caption{\textbf{Directional-dependent magnetization of La$_2$NiO$_{4-x}$F$_{2x}$ single crystals.} For comparison, data from as-grown La$_2$NiO$_{4.18}$ (red) and reduced La$_2$NiO$_4$ (black) are put as a top panel. The inset shows a magnification of the high temperature transition in reduced La$_2$NiO$_4$ (black). The bottom panel shows the temperature-dependent magnetization measured in a field of 0.1 T along the $a$-, $b-$ and $c$-axes for a twice PTFE treated crystal. Pronounced anisotropy is observed in the $a$-axis ZFC curve (black) compared to $c$ (red). The 80 K transition is put as a reference data of NiF$_2$~\cite{Arumugam2019} (green hollow spheres)}
    \label{chi_fig}
\end{figure}

We present a comprehensive study of the magnetic properties of single-crystalline La$_2$NiO$_{4-x}$F$_{2x}$, focusing on the interplay between fluorination, magnetic anisotropy, and potential phase impurities. Zero-field-cooled (ZFC) and field-cooled (FC) magnetization measurements were performed in a magnetic field of 0.1 T along the $a$-axis over a broad temperature range (2--800 K), with results displayed in Fig.~\ref{chi_fig}(a) as blue (La$_2$NiO$_{4.18}$) and red (reduced La$_2$NiO$_4$) curves for comparison. The pristine oxygen-intercalated La$_2$NiO$_{4.18}$ exhibits a single, well-defined antiferromagnetic (AFM) transition near 22 K, while the reduced La$_2$NiO$_4$ shows a sharp increase in magnetization at $\approx$646 K indicative of the well-known high-temperature AFM ordering at 650 K~\cite{Lander1989}, followed by a continuous decrease, characteristic of antiferromagnetic behavior. Below this transition, the magnetization exhibits strong history dependence, reflecting the presence of residual magnetic order and potential inhomogeneities. Additional features are observed at $\approx$350 K and 80 K, suggesting secondary magnetic transitions or spin reorientation phenomena.

To assess possible impurity contributions, we include the magnetization data of NiF$_2$ from Ref.~\cite{Arumugam2019}, which displays a sharp AFM transition at 68.5 K. Although X-ray diffraction (XRD) reveals no bulk NiF$_2$ phase, energy-dispersive X-ray spectroscopy (EDX) indicates surface overfluorination, raising concerns about local chemical inhomogeneity. Given the striking similarity in transition temperature between NiF$_2$ and the 80 K feature reported in powder samples \cite{wissel_LNOF}, a careful evaluation of impurity effects is warranted.

The magnetic behavior of La$_2$NiO$_{4-x}$F$_{2x}$ was previously studied in powder form~\cite{wissel_LNOF,Jacobs2024}, where a paramagnetic-to-AFM transition was observed at 49 K, with no ZFC--FC splitting and a Curie--Weiss fit yielding a paramagnetic moment of 2.82~$\mu_\mathrm{B}$, in agreement with the spin-only value for high-spin Ni$^{2+}$ (2.83~$\mu_\mathrm{B}$). A large negative Weiss constant ($\theta \approx -648$~K) confirms strong AFM interactions. Considering the high magnetic transition temperature of La$_2$NiO$_4$ this is reasonable. A weak ferromagnetic signal below 30 K was attributed to magnetic canting within the AFM lattice.

In our single-crystal study, ZFC and FC magnetization along the $c$-axis (grey curve in Fig.~\ref{chi_fig}(a)) closely resemble the powder data: a clear AFM transition near 50 K, absence of ZFC--FC splitting, and linear inverse susceptibility between 150 and 300 K, consistent with Curie--Weiss behavior. However, the paramagnetic moment is elevated to 3.40(2)~$\mu_\mathrm{B}$, significantly above the expected Ni$^{2+}$ value, suggesting either enhanced orbital contributions or the presence of additional magnetic phases. The Weiss constant is also markedly large ($\theta \approx -805(10)$~K), indicating strong AFM correlations. Notably, the low-temperature rise in magnetization below 30 K, observed in the powder sample, is absent in our data.

In comparison, measurements along the $a$-axis (black dashed curve) reveal pronounced magnetic anisotropy. The ZFC curve exhibits a sharper, more distinct transition at $\approx$50 K, while the FC curve remains nearly constant, suggesting that the magnetic moments are preferentially aligned within the $ab$-plane. This anisotropy is further highlighted by the high-temperature behavior: a rising background in the $a$-axis data above 420 K, indicative of sample decomposition, suggests the presence of unreacted La$_2$NiO$_4$, which itself exhibits a sharp AFM transition at 80 K and a secondary transition near 350 K. A Curie--Weiss fit along the $a$-axis yields a more reasonable moment of 2.87(2)~$\mu_\mathrm{B}$, but the large negative Weiss constant persists. Considering the superstructural solution of possibly intercalated fluorine the curve looks quite like a combination of a behavior of oxygenated La$_2$NiO$_{4+\delta}$ while delta here is rather intercalated F with partial NiF$_2$ contribution. 

To further probe the origin of the observed features, we extended measurements up to 800 K using an oven-equipped cryostat. The curve looks remarkably similar to oxygen intercalated nickleats. The combined influence of unfluorinated La$_2$NiO$_4$ (with transitions at 80 K and 350 K) and surface NiF$_2$ (at 68.5 K) could mimic the observed magnetization profile. However, the persistence of a well-defined transition with anisotropy near 50 K across multiple samples, especially in the absence of bulk crystalline NiF$_2$, suggests the emergence of a unique, intrinsic magnetic order of surface La$_2$NiO$_{4-x}$F$_{2x}$.

In summary, our data reveal a rich magnetic landscape in fluorinated La$_2$NiO$_4$, characterized by strong anisotropy, a well-defined AFM transition near 50 K, and a intercalated La$_2$NiO$_{4}$F$_x$. While impurities cannot be entirely ruled out, the consistency of the low-temperature transition with fluorination levels and the absence of bulk NiF$_2$ indicates that the observed magnetism is intrinsic to the highly fluorinated surface La$_2$NiO$_{4-x}$F$_{2x}$, but might be equally explained by diluted NiF$_2$.

\section{Conclusions}
Single-crystalline $n$ = 1 Ruddlesden–-Popper  La$_2$NiO$_{4+\delta}$ was successfully grown by the optical floating zone (OFZ) method and confirmed to crystallize the tetragonal \textit{I}4/\textit{mmm} structure by both powder and single-crystal X-ray diffraction. TGA studies reveal a stoichiometry La$_2$NiO$_{4.15(3)}$ in the as grown boule, resulting in a full reduction of the bulk crystal confirmed by PXRD to be the established \textit{Bmab}. Topochemical fluorination using PTFE, PVDF, and CuF$_2$ was systematically investigated to deeply understand fluorine incorporation mechanisms in single crystals. Among the fluorination agents, PTFE was found to be the most effective, resulting in a symmetry lowering from tetragonal \textit{I}4/\textit{mmm} to a complex superstructure with increased unit cell volume compared to the tetragonal aristotype and $C$2/$m$ symmetry. The symmetry lowering originiates from a channel like ordering of interstitial sites within the laers, which has not been observed before for anion-intercalated RP-type compounds. Unlike reduction the crystal quality remains excellent. Comparative EDX analyses revealed that fluorine uptake strongly depends on both fluorine source and thermal conditions, with higher temperatures increasing diffusion-driven incorporation for PVDF and CuF$_2$. Successive PTFE treatments resulted in significantly higher fluorine contents, indicating a diffusion-limited process with increasing penetration depth. However, cross-sectional EDX mapping revealed notable fluorine inhomogeneity, characterized by strong surface accumulation and significant concentration gradients throughout the bulk crystal. These results highlight both the effectiveness and the intrinsic limitations of topochemical fluorination in RP-type single crystals, emphasizing that notable non-uniform fluorine incorporation must be carefully considered when interpreting the structural and physical properties of fluorinated nickelates single crystals. Considering the very thin surface fluorination layer grain size considerations for powder are extremely relevant. The observation of F in EDX and a stoichiometry of O$_{3.9(3)}$ indicate that the superstructure stems from F and not oxygen intercalation. Magnetic characterization show a magnetic anisotrop sample supporting single crystallinity, we find that the magnetic transition around 50 K can also be explained by surface NiF$_2$, while the bulk shows an intercalated La$_2$NiO$_{4}$F$_i$ like behavior consistent with EDX and single crystal XRD results. Overall, the result gave strong insight into the fluorination process for all specimen. Recent development in the nickelate superconductor field also shows our work will be of considerable importance, as intercalated oxygen is considerable in ozone annealed films \cite{FerencSegedin2026}, which is required to obtain superconductivity \cite{Ko2024}. Via fluorine intercalation a NMR tracable variant exists without destruction of the crystallinity.

\section{Methods}
\subsection{Synthesis and Topochemical Fluorination of La\texorpdfstring{$_2$NiO$_{4+\delta}$}{2NiO4+delta} Single Crystals}
\subsubsection{Crystal Growth}

La$_2$NiO$_{4+\delta}$ single crystals were synthesized using the Optical Float Zone (OFZ) method. Before synthesis, La$_2$O$_3$ (Alfa Aesar, 99.9\%) was dried at 1100\degree C for 12~h, while NiO (Alfa Aesar, 99.0\%) was dried at 700\degree C for 6~h. The stoichiometric mixtures of La$_2$O$_3$ and NiO were ball-milled at 300~rpm for 6~h and then sintered twice in air at 1200\degree C for 12~h with intermediate grinding. The as-sintered powder was formed into cylindrical rods (\o=4~mm and $l$=100~mm) using an isostatic press (70~MPa) with rubber forms and subsequently annealed at 1200\degree C for 12~h in air. Single crystal growth was carried out in an Optical Float Zone Furnace (FZ-T-10000-H-III-VPR) using four 1000 W halogen lamps. During growth, the feed and seed rods were counter-rotated at 24 rpm to minimize the diffusion zone near the solid–liquid interface. Argon flow of 100 cc~min$^{-1}$ was applied, and the growth rate stabilized at 4~mm~h$^{-1}$.\\
\subsubsection{Topochemical Fluorination of La$_2$NiO$_{4+\delta}$}
\begin{figure}[h]
\centering
\includegraphics[width=1.0\columnwidth]{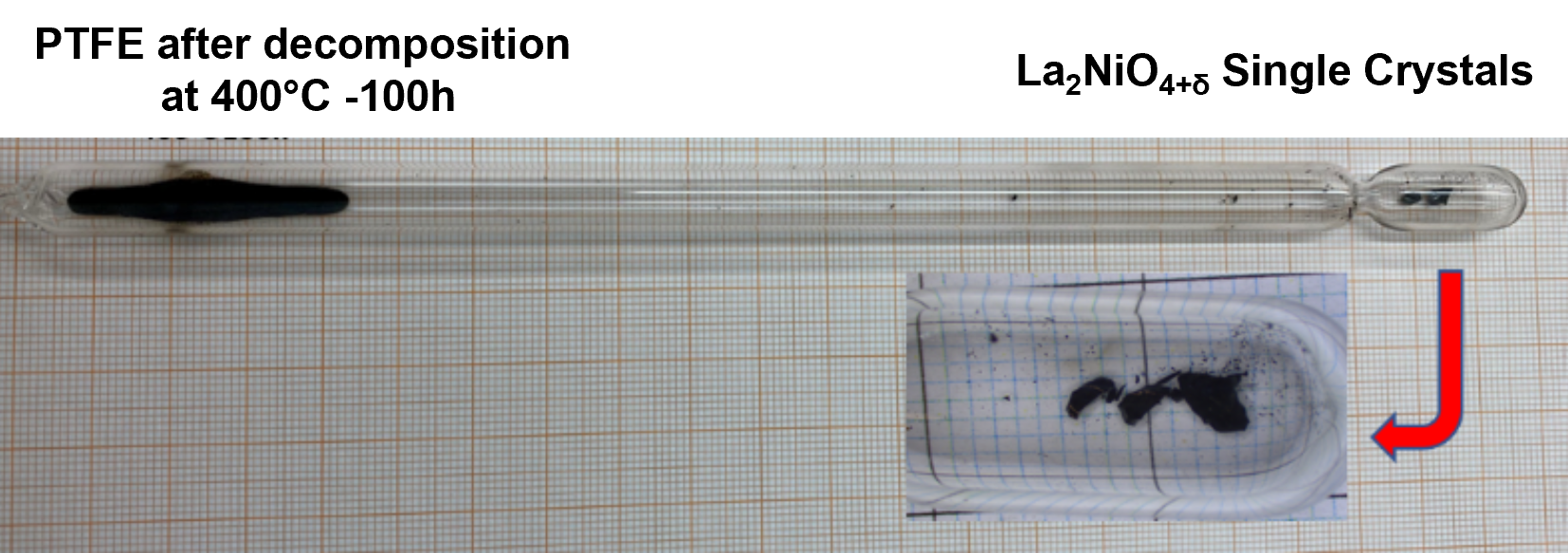}
\caption{Prepared glass reactors after fluorination process with PTFE
}
\label{quartzampoule}
\end{figure}

Topochemical fluorination of La$_2$NiO$_{4+\delta}$ single crystals was carried out using different fluorination agents. Polymer-based PTFE (polytetrafluoroethylene, (C$_2$F$_4$)$_n$, Sigma-Aldrich) and PVDF (polyvinylidene fluoride, (C$_2$H$_2$F$_2$)$_n$, Sigma-Aldrich), as well as the inorganic fluorination agent CuF$_2$ (copper(II) fluoride dihydrate, abcr GmbH), were tested at temperatures corresponding to their decomposition points. Both direct and indirect contact between the crystals and the fluorination agents were examined. For the experiments, glass reactors (ampoule) were prepared containing La$_2$NiO$_{4+\delta}$ single crystals together with the fluorination agents, enabling gas-phase diffusion of fluorine released during decomposition. In the direct-contact method, as-grown single crystals (sizes between 300~\micro m and 1~mm, total mass $\approx$~100~mg) were embedded in the fluorination agents. In the indirect-contact method, the crystals and fluorination agents were placed at opposite ends of the reactor (see Fig.~\ref{quartzampoule}). The mass ratio between La$_2$NiO$_{4+\delta}$ and the fluorination agents was adjusted according to the type of agent: a 1:5 ratio was used for PTFE and PVDF, while a 1:3 ratio was applied for CuF$_2$.\\
Before the fluorination process, the glass reactors were evacuated to a pressure of 10$^{-6}$~mbar and sealed. The sealed reactors were placed in a tube furnace and heated for 100~h under the following conditions: 400\degree C for PVDF, 400\degree C for PTFE, and 250–400\degree C for CuF$_2$. After fluorination process, the glass reactors were opened in a fume hood to release residual gases. This procedure was repeated twice using fresh fluorination agents to ensure sufficient reaction. The resulting fluorinated crystals were collected and prepared for subsequent analyses.
\subsection{X-ray Diffraction Measurements}
The crystallinity and phase purity of the as-grown single crystal were examined using X-ray diffraction on powdered samples. Powder XRD measurements were performed on ground crystals to identify possible impurity phases and to determine the crystal structure and lattice parameters. The diffraction patterns were recorded on a Rigaku SmartLab in Bragg–Brentano geometry, equipped with a Cu-K$_{\alpha}$radiation ($\lambda$~=~1.5406~\AA) source operating at 40~kV and 30~mA and a HyPix-3000 detector. For powder samples, data were obtained the 10\degree – 80\degree~$2\theta$ range with a step size of 0.005\degree and a fixed divergence slit of 0.3\degree. Rietveld refinements were carried out using TOPAS V6 software. The instrumental intensity distribution was determined empirically from a fundamental parameter set, based on a reference scan of LaB$_6$ (NIST 660a). Microstructural parameters such as crystallite size and strain were refined to account for sample-dependent peak broadening.\\
The single-crystal XRD data was collected at room temperature with a Rigaku XtaLab Mini II Single Crystal X-ray Diffractometer using Mo-K$_{\alpha}$ radiation ($\lambda$~=~0.71073~\AA), graphite monochromator, operated 50~kV and 12~mA 600~W. All the crystals analyzed in this study are fragments of larger crystals broken off in ethanol since very small pieces of crystals are more suited for analysis by single-crystal XRD. The pieces were mounted with some grease on the top of capillary tube. Data was integrated with CrysAlisPro 1.171.42.54a (Rigaku Oxford Diffraction, 2022) and refinements were performed with Olex 2-1.5 Software.
\subsection{Scanning Electron Microscopy}
The SEM images were taken using Zeiss DSM 982 Gemini SEM equipped with SE and BSE detectors by operating at 12~keV. Energy-dispersive X-ray spectra were taken using a NORAN System 7 (NSS212E) detector in a Tescan Vega (TS-5130MM) SEM. The samples were first cleaved to reveal the inner cross-section and then embedded in conductive carbon paste prior to measurement.
\subsection{Magnetic Susceptibility Measurements}
Magnetic susceptibility measurements were performed using a vibrating sample magnetometer (MPMS 3, Quantum Design) from 2-300 K and from 300-800 K using the oven option. 

\section*{Data availability}
The data that support the findings of this study are available from Pascal Puphal (puphal@fkf.mpg.de) upon reasonable request.

\newpage
\bibliography{Literature}

\begin{thebibliography}{39}%
\makeatletter
\providecommand \@ifxundefined [1]{%
 \@ifx{#1\undefined}
}%
\providecommand \@ifnum [1]{%
 \ifnum #1\expandafter \@firstoftwo
 \else \expandafter \@secondoftwo
 \fi
}%
\providecommand \@ifx [1]{%
 \ifx #1\expandafter \@firstoftwo
 \else \expandafter \@secondoftwo
 \fi
}%
\providecommand \natexlab [1]{#1}%
\providecommand \enquote  [1]{``#1''}%
\providecommand \bibnamefont  [1]{#1}%
\providecommand \bibfnamefont [1]{#1}%
\providecommand \citenamefont [1]{#1}%
\providecommand \href@noop [0]{\@secondoftwo}%
\providecommand \href [0]{\begingroup \@sanitize@url \@href}%
\providecommand \@href[1]{\@@startlink{#1}\@@href}%
\providecommand \@@href[1]{\endgroup#1\@@endlink}%
\providecommand \@sanitize@url [0]{\catcode `\\12\catcode `\$12\catcode `\&12\catcode `\#12\catcode `\^12\catcode `\_12\catcode `\%12\relax}%
\providecommand \@@startlink[1]{}%
\providecommand \@@endlink[0]{}%
\providecommand \url  [0]{\begingroup\@sanitize@url \@url }%
\providecommand \@url [1]{\endgroup\@href {#1}{\urlprefix }}%
\providecommand \urlprefix  [0]{URL }%
\providecommand \Eprint [0]{\href }%
\providecommand \doibase [0]{https://doi.org/}%
\providecommand \selectlanguage [0]{\@gobble}%
\providecommand \bibinfo  [0]{\@secondoftwo}%
\providecommand \bibfield  [0]{\@secondoftwo}%
\providecommand \translation [1]{[#1]}%
\providecommand \BibitemOpen [0]{}%
\providecommand \bibitemStop [0]{}%
\providecommand \bibitemNoStop [0]{.\EOS\space}%
\providecommand \EOS [0]{\spacefactor3000\relax}%
\providecommand \BibitemShut  [1]{\csname bibitem#1\endcsname}%
\let\auto@bib@innerbib\@empty
\bibitem [{\citenamefont {Yang}\ \emph {et~al.}(2025)\citenamefont {Yang}, \citenamefont {Ortiz}, \citenamefont {Wang}, \citenamefont {Sigle}, \citenamefont {Anggara}, \citenamefont {Benckiser}, \citenamefont {Keimer},\ and\ \citenamefont {van Aken}}]{Yang2025_ncomm_3277}%
  \BibitemOpen
  \bibfield  {author} {\bibinfo {author} {\bibfnamefont {C.}~\bibnamefont {Yang}}, \bibinfo {author} {\bibfnamefont {R.~A.}\ \bibnamefont {Ortiz}}, \bibinfo {author} {\bibfnamefont {H.}~\bibnamefont {Wang}}, \bibinfo {author} {\bibfnamefont {W.}~\bibnamefont {Sigle}}, \bibinfo {author} {\bibfnamefont {K.}~\bibnamefont {Anggara}}, \bibinfo {author} {\bibfnamefont {E.}~\bibnamefont {Benckiser}}, \bibinfo {author} {\bibfnamefont {B.}~\bibnamefont {Keimer}},\ and\ \bibinfo {author} {\bibfnamefont {P.~A.}\ \bibnamefont {van Aken}},\ }\bibfield  {title} {\bibinfo {title} {Atomic-scale observation of geometric reconstruction in a fluorine-intercalated infinite layer nickelate superlattice},\ }\bibfield  {journal} {\bibinfo  {journal} {Nature Communications}\ }\textbf {\bibinfo {volume} {16}},\ \href {https://doi.org/10.1038/s41467-025-58646-0} {10.1038/s41467-025-58646-0} (\bibinfo {year} {2025})\BibitemShut {NoStop}%
\bibitem [{\citenamefont {Zhang}\ \emph {et~al.}(2016)\citenamefont {Zhang}, \citenamefont {Senn},\ and\ \citenamefont {Hayward}}]{Zhang2016}%
  \BibitemOpen
  \bibfield  {author} {\bibinfo {author} {\bibfnamefont {R.}~\bibnamefont {Zhang}}, \bibinfo {author} {\bibfnamefont {M.~S.}\ \bibnamefont {Senn}},\ and\ \bibinfo {author} {\bibfnamefont {M.~A.}\ \bibnamefont {Hayward}},\ }\bibfield  {title} {\bibinfo {title} {Directed lifting of inversion symmetry in ruddlesden–popper oxide–fluorides: Toward ferroelectric and multiferroic behavior},\ }\href {https://doi.org/10.1021/acs.chemmater.6b03931} {\bibfield  {journal} {\bibinfo  {journal} {Chemistry of Materials}\ }\textbf {\bibinfo {volume} {28}},\ \bibinfo {pages} {8399} (\bibinfo {year} {2016})}\BibitemShut {NoStop}%
\bibitem [{\citenamefont {Herlihy}\ \emph {et~al.}(2025)\citenamefont {Herlihy}, \citenamefont {Chen}, \citenamefont {Ritter}, \citenamefont {Chuang},\ and\ \citenamefont {Senn}}]{Herlihy2025}%
  \BibitemOpen
  \bibfield  {author} {\bibinfo {author} {\bibfnamefont {A.}~\bibnamefont {Herlihy}}, \bibinfo {author} {\bibfnamefont {W.-T.}\ \bibnamefont {Chen}}, \bibinfo {author} {\bibfnamefont {C.}~\bibnamefont {Ritter}}, \bibinfo {author} {\bibfnamefont {Y.-C.}\ \bibnamefont {Chuang}},\ and\ \bibinfo {author} {\bibfnamefont {M.~S.}\ \bibnamefont {Senn}},\ }\bibfield  {title} {\bibinfo {title} {Interplay between jahn–teller distortions and structural phase transitions in ruddlesden--poppers},\ }\href {https://doi.org/10.1021/jacs.5c00459} {\bibfield  {journal} {\bibinfo  {journal} {Journal of the American Chemical Society}\ }\textbf {\bibinfo {volume} {147}},\ \bibinfo {pages} {7209} (\bibinfo {year} {2025})},\ \bibinfo {note} {pMID: 39964199},\ \Eprint {https://arxiv.org/abs/https://doi.org/10.1021/jacs.5c00459} {https://doi.org/10.1021/jacs.5c00459} \BibitemShut {NoStop}%
\bibitem [{\citenamefont {Flathmann}\ \emph {et~al.}(2024)\citenamefont {Flathmann}, \citenamefont {Meyer}, \citenamefont {Ross}, \citenamefont {Dehning}, \citenamefont {Jooss},\ and\ \citenamefont {Seibt}}]{Flathmann2024_APLMat_061112}%
  \BibitemOpen
  \bibfield  {author} {\bibinfo {author} {\bibfnamefont {C.}~\bibnamefont {Flathmann}}, \bibinfo {author} {\bibfnamefont {T.}~\bibnamefont {Meyer}}, \bibinfo {author} {\bibfnamefont {U.}~\bibnamefont {Ross}}, \bibinfo {author} {\bibfnamefont {A.}~\bibnamefont {Dehning}}, \bibinfo {author} {\bibfnamefont {C.}~\bibnamefont {Jooss}},\ and\ \bibinfo {author} {\bibfnamefont {M.}~\bibnamefont {Seibt}},\ }\bibfield  {title} {\bibinfo {title} {Relationship between structure and charge/orbital order in epitaxial single layer ruddlesden-popper manganite thin films},\ }\href {https://doi.org/10.1063/5.0208123} {\bibfield  {journal} {\bibinfo  {journal} {APL Materials}\ }\textbf {\bibinfo {volume} {12}},\ \bibinfo {pages} {061112} (\bibinfo {year} {2024})}\BibitemShut {NoStop}%
\bibitem [{\citenamefont {Balachandran}\ \emph {et~al.}(2014)\citenamefont {Balachandran}, \citenamefont {Puggioni},\ and\ \citenamefont {Rondinelli}}]{Balachandran2014}%
  \BibitemOpen
  \bibfield  {author} {\bibinfo {author} {\bibfnamefont {P.~V.}\ \bibnamefont {Balachandran}}, \bibinfo {author} {\bibfnamefont {D.}~\bibnamefont {Puggioni}},\ and\ \bibinfo {author} {\bibfnamefont {J.~M.}\ \bibnamefont {Rondinelli}},\ }\bibfield  {title} {\bibinfo {title} {Crystal-chemistry guidelines for noncentrosymmetric $\mathrm{A}_{2}\mathrm{B}\mathrm{O}_{4}$ ruddlesden--popper oxides},\ }\href {https://doi.org/10.1021/ic402283c} {\bibfield  {journal} {\bibinfo  {journal} {Inorganic Chemistry}\ }\textbf {\bibinfo {volume} {53}},\ \bibinfo {pages} {336} (\bibinfo {year} {2014})},\ \Eprint {https://arxiv.org/abs/https://doi.org/10.1021/ic402283c} {https://doi.org/10.1021/ic402283c} \BibitemShut {NoStop}%
\bibitem [{\citenamefont {Vanita}\ \emph {et~al.}(2024)\citenamefont {Vanita}, \citenamefont {Waidha}, \citenamefont {Vasala}, \citenamefont {Puphal}, \citenamefont {Schoch}, \citenamefont {Glatzel}, \citenamefont {Bauer},\ and\ \citenamefont {Clemens}}]{Vanita2024}%
  \BibitemOpen
  \bibfield  {author} {\bibinfo {author} {\bibfnamefont {V.}~\bibnamefont {Vanita}}, \bibinfo {author} {\bibfnamefont {A.~I.}\ \bibnamefont {Waidha}}, \bibinfo {author} {\bibfnamefont {S.}~\bibnamefont {Vasala}}, \bibinfo {author} {\bibfnamefont {P.}~\bibnamefont {Puphal}}, \bibinfo {author} {\bibfnamefont {R.}~\bibnamefont {Schoch}}, \bibinfo {author} {\bibfnamefont {P.}~\bibnamefont {Glatzel}}, \bibinfo {author} {\bibfnamefont {M.}~\bibnamefont {Bauer}},\ and\ \bibinfo {author} {\bibfnamefont {O.}~\bibnamefont {Clemens}},\ }\bibfield  {title} {\bibinfo {title} {Insights into the first multi-transition-metal containing ruddlesden–popper-type cathode for all-solid-state fluoride ion batteries},\ }\href {https://doi.org/10.1039/D4TA00704B} {\bibfield  {journal} {\bibinfo  {journal} {J. Mater. Chem. A}\ }\textbf {\bibinfo {volume} {12}},\ \bibinfo {pages} {8769} (\bibinfo {year} {2024})}\BibitemShut {NoStop}%
\bibitem [{\citenamefont {Vanita}\ \emph {et~al.}(2025)\citenamefont {Vanita}, \citenamefont {Schoch}, \citenamefont {Puphal}, \citenamefont {Yilmaz}, \citenamefont {Bauer},\ and\ \citenamefont {Clemens}}]{Vanita2025}%
  \BibitemOpen
  \bibfield  {author} {\bibinfo {author} {\bibfnamefont {V.}~\bibnamefont {Vanita}}, \bibinfo {author} {\bibfnamefont {R.}~\bibnamefont {Schoch}}, \bibinfo {author} {\bibfnamefont {P.}~\bibnamefont {Puphal}}, \bibinfo {author} {\bibfnamefont {H.}~\bibnamefont {Yilmaz}}, \bibinfo {author} {\bibfnamefont {M.}~\bibnamefont {Bauer}},\ and\ \bibinfo {author} {\bibfnamefont {O.}~\bibnamefont {Clemens}},\ }\bibfield  {title} {\bibinfo {title} {Structural and electrochemical behaviour of bilayer manganite $\mathrm{LaSr}_{2}\mathrm{Mn}_{2}\mathrm{O}_{6.96}$ cathode for all-solid-state fluoride ion batteries},\ }\href {https://doi.org/https://doi.org/10.1016/j.actphy.2025.100181} {\bibfield  {journal} {\bibinfo  {journal} {Acta Physico-Chimica Sinica}\ ,\ \bibinfo {pages} {100181}} (\bibinfo {year} {2025})}\BibitemShut {NoStop}%
\bibitem [{\citenamefont {Yilmaz}\ \emph {et~al.}(2025)\citenamefont {Yilmaz}, \citenamefont {Sosa-Lizama}, \citenamefont {Knauft}, \citenamefont {Küster}, \citenamefont {Starke}, \citenamefont {Isobe}, \citenamefont {Clemens}, \citenamefont {van Aken}, \citenamefont {Suyolcu},\ and\ \citenamefont {Puphal}}]{Yilmaz2025_CommPhys_408}%
  \BibitemOpen
  \bibfield  {author} {\bibinfo {author} {\bibfnamefont {H.}~\bibnamefont {Yilmaz}}, \bibinfo {author} {\bibfnamefont {P.}~\bibnamefont {Sosa-Lizama}}, \bibinfo {author} {\bibfnamefont {M.}~\bibnamefont {Knauft}}, \bibinfo {author} {\bibfnamefont {K.}~\bibnamefont {Küster}}, \bibinfo {author} {\bibfnamefont {U.}~\bibnamefont {Starke}}, \bibinfo {author} {\bibfnamefont {M.}~\bibnamefont {Isobe}}, \bibinfo {author} {\bibfnamefont {O.}~\bibnamefont {Clemens}}, \bibinfo {author} {\bibfnamefont {P.~A.}\ \bibnamefont {van Aken}}, \bibinfo {author} {\bibfnamefont {Y.~E.}\ \bibnamefont {Suyolcu}},\ and\ \bibinfo {author} {\bibfnamefont {P.}~\bibnamefont {Puphal}},\ }\bibfield  {title} {\bibinfo {title} {Floating zone growth of large tetragonal ruddlesden--popper bilayer nickelate $\mathrm{Y}_{y}\mathrm{Sr}_{3-y}\mathrm{Ni}_{2-x}\mathrm{Al}_{x}\mathrm{O}_{7-\delta}$ single crystals},\ }\href {https://doi.org/10.1038/s42005-025-02340-6} {\bibfield  {journal} {\bibinfo  {journal} {Communications Physics}\ }\textbf
  {\bibinfo {volume} {8}},\ \bibinfo {pages} {408} (\bibinfo {year} {2025})}\BibitemShut {NoStop}%
\bibitem [{\citenamefont {Wissel}\ \emph {et~al.}(2020{\natexlab{a}})\citenamefont {Wissel}, \citenamefont {Vogel}, \citenamefont {Dasgupta}, \citenamefont {Fortes}, \citenamefont {Slater},\ and\ \citenamefont {Clemens}}]{Wissel2020}%
  \BibitemOpen
  \bibfield  {author} {\bibinfo {author} {\bibfnamefont {K.}~\bibnamefont {Wissel}}, \bibinfo {author} {\bibfnamefont {T.}~\bibnamefont {Vogel}}, \bibinfo {author} {\bibfnamefont {S.}~\bibnamefont {Dasgupta}}, \bibinfo {author} {\bibfnamefont {A.~D.}\ \bibnamefont {Fortes}}, \bibinfo {author} {\bibfnamefont {P.~R.}\ \bibnamefont {Slater}},\ and\ \bibinfo {author} {\bibfnamefont {O.}~\bibnamefont {Clemens}},\ }\bibfield  {title} {\bibinfo {title} {Topochemical fluorination of $n$ = 2 ruddlesden–popper type $\mathrm{Sr}_{3}\mathrm{Ti}_{2}\mathrm{O}_{7}$ to $\mathrm{Sr}_{3}\mathrm{Ti}_{2}\mathrm{O}_{5}\mathrm{F}_{4}$ and its reductive defluorination},\ }\href {https://doi.org/10.1021/acs.inorgchem.9b02783} {\bibfield  {journal} {\bibinfo  {journal} {Inorganic Chemistry}\ }\textbf {\bibinfo {volume} {59}},\ \bibinfo {pages} {1153} (\bibinfo {year} {2020}{\natexlab{a}})},\ \bibinfo {note} {pMID: 31880431},\ \Eprint {https://arxiv.org/abs/https://doi.org/10.1021/acs.inorgchem.9b02783}
  {https://doi.org/10.1021/acs.inorgchem.9b02783} \BibitemShut {NoStop}%
\bibitem [{\citenamefont {Wissel}\ \emph {et~al.}(2018)\citenamefont {Wissel}, \citenamefont {Heldt}, \citenamefont {Groszewicz}, \citenamefont {Dasgupta}, \citenamefont {Breitzke}, \citenamefont {Donzelli}, \citenamefont {Waidha}, \citenamefont {Fortes}, \citenamefont {Rohrer}, \citenamefont {Slater}, \citenamefont {Buntkowsky},\ and\ \citenamefont {Clemens}}]{wissel_LNOF}%
  \BibitemOpen
  \bibfield  {author} {\bibinfo {author} {\bibfnamefont {K.}~\bibnamefont {Wissel}}, \bibinfo {author} {\bibfnamefont {J.}~\bibnamefont {Heldt}}, \bibinfo {author} {\bibfnamefont {P.~B.}\ \bibnamefont {Groszewicz}}, \bibinfo {author} {\bibfnamefont {S.}~\bibnamefont {Dasgupta}}, \bibinfo {author} {\bibfnamefont {H.}~\bibnamefont {Breitzke}}, \bibinfo {author} {\bibfnamefont {M.}~\bibnamefont {Donzelli}}, \bibinfo {author} {\bibfnamefont {A.~I.}\ \bibnamefont {Waidha}}, \bibinfo {author} {\bibfnamefont {A.~D.}\ \bibnamefont {Fortes}}, \bibinfo {author} {\bibfnamefont {J.}~\bibnamefont {Rohrer}}, \bibinfo {author} {\bibfnamefont {P.~R.}\ \bibnamefont {Slater}}, \bibinfo {author} {\bibfnamefont {G.}~\bibnamefont {Buntkowsky}},\ and\ \bibinfo {author} {\bibfnamefont {O.}~\bibnamefont {Clemens}},\ }\bibfield  {title} {\bibinfo {title} {Topochemical fluorination of $\mathrm{La}_{2}\mathrm{Ni}\mathrm{O}_{4+\delta}$: Unprecedented ordering of oxide and fluoride ions in
  $\mathrm{La}_{2}\mathrm{Ni}\mathrm{O}_{3}\mathrm{F}_{2}$},\ }\href {https://doi.org/10.1021/acs.inorgchem.8b00661} {\bibfield  {journal} {\bibinfo  {journal} {Inorganic Chemistry}\ }\textbf {\bibinfo {volume} {57}},\ \bibinfo {pages} {6549} (\bibinfo {year} {2018})},\ \bibinfo {note} {pMID: 29749739}\BibitemShut {NoStop}%
\bibitem [{\citenamefont {Puphal}\ \emph {et~al.}(2023{\natexlab{a}})\citenamefont {Puphal}, \citenamefont {Wehinger}, \citenamefont {Nuss}, \citenamefont {K\"uster}, \citenamefont {Starke}, \citenamefont {Garbarino}, \citenamefont {Keimer}, \citenamefont {Isobe},\ and\ \citenamefont {Hepting}}]{Puphal_LNO2}%
  \BibitemOpen
  \bibfield  {author} {\bibinfo {author} {\bibfnamefont {P.}~\bibnamefont {Puphal}}, \bibinfo {author} {\bibfnamefont {B.}~\bibnamefont {Wehinger}}, \bibinfo {author} {\bibfnamefont {J.}~\bibnamefont {Nuss}}, \bibinfo {author} {\bibfnamefont {K.}~\bibnamefont {K\"uster}}, \bibinfo {author} {\bibfnamefont {U.}~\bibnamefont {Starke}}, \bibinfo {author} {\bibfnamefont {G.}~\bibnamefont {Garbarino}}, \bibinfo {author} {\bibfnamefont {B.}~\bibnamefont {Keimer}}, \bibinfo {author} {\bibfnamefont {M.}~\bibnamefont {Isobe}},\ and\ \bibinfo {author} {\bibfnamefont {M.}~\bibnamefont {Hepting}},\ }\bibfield  {title} {\bibinfo {title} {Synthesis and physical properties of $\mathrm{LaNiO}_{2}$ crystals},\ }\href {https://doi.org/10.1103/PhysRevMaterials.7.014804} {\bibfield  {journal} {\bibinfo  {journal} {Phys. Rev. Mater.}\ }\textbf {\bibinfo {volume} {7}},\ \bibinfo {pages} {014804} (\bibinfo {year} {2023}{\natexlab{a}})}\BibitemShut {NoStop}%
\bibitem [{\citenamefont {Puphal}\ \emph {et~al.}(2025)\citenamefont {Puphal}, \citenamefont {Schäfer}, \citenamefont {Keimer},\ and\ \citenamefont {Hepting}}]{Puphal2025_NatRevPhys}%
  \BibitemOpen
  \bibfield  {author} {\bibinfo {author} {\bibfnamefont {P.}~\bibnamefont {Puphal}}, \bibinfo {author} {\bibfnamefont {T.}~\bibnamefont {Schäfer}}, \bibinfo {author} {\bibfnamefont {B.}~\bibnamefont {Keimer}},\ and\ \bibinfo {author} {\bibfnamefont {M.}~\bibnamefont {Hepting}},\ }\bibfield  {title} {\bibinfo {title} {Superconductivity in infinite-layer and ruddlesden--popper nickelates},\ }\bibfield  {journal} {\bibinfo  {journal} {Nature Reviews Physics}\ }\href {https://doi.org/10.1038/s42254-025-00898-2} {10.1038/s42254-025-00898-2} (\bibinfo {year} {2025})\BibitemShut {NoStop}%
\bibitem [{\citenamefont {Peterson}\ \emph {et~al.}(2018)\citenamefont {Peterson}, \citenamefont {Swift}, \citenamefont {Porter}, \citenamefont {Cl\'ement}, \citenamefont {Wu}, \citenamefont {Ahn}, \citenamefont {Moon}, \citenamefont {Chakoumakos}, \citenamefont {Ruff}, \citenamefont {Cao}, \citenamefont {Van~de Walle},\ and\ \citenamefont {Wilson}}]{Peterson_Sr3Ir2O7F2}%
  \BibitemOpen
  \bibfield  {author} {\bibinfo {author} {\bibfnamefont {C.}~\bibnamefont {Peterson}}, \bibinfo {author} {\bibfnamefont {M.~W.}\ \bibnamefont {Swift}}, \bibinfo {author} {\bibfnamefont {Z.}~\bibnamefont {Porter}}, \bibinfo {author} {\bibfnamefont {R.~J.}\ \bibnamefont {Cl\'ement}}, \bibinfo {author} {\bibfnamefont {G.}~\bibnamefont {Wu}}, \bibinfo {author} {\bibfnamefont {G.~H.}\ \bibnamefont {Ahn}}, \bibinfo {author} {\bibfnamefont {S.~J.}\ \bibnamefont {Moon}}, \bibinfo {author} {\bibfnamefont {B.~C.}\ \bibnamefont {Chakoumakos}}, \bibinfo {author} {\bibfnamefont {J.~P.~C.}\ \bibnamefont {Ruff}}, \bibinfo {author} {\bibfnamefont {H.}~\bibnamefont {Cao}}, \bibinfo {author} {\bibfnamefont {C.}~\bibnamefont {Van~de Walle}},\ and\ \bibinfo {author} {\bibfnamefont {S.~D.}\ \bibnamefont {Wilson}},\ }\bibfield  {title} {\bibinfo {title} {$\mathrm{Sr}_{3}\mathrm{Ir}_{2}\mathrm{O}_{7}\mathrm{F}_{2}$: Topochemical conversion of a relativistic mott state into a spin-orbit driven band insulator},\ }\href
  {https://doi.org/10.1103/PhysRevB.98.155128} {\bibfield  {journal} {\bibinfo  {journal} {Phys. Rev. B}\ }\textbf {\bibinfo {volume} {98}},\ \bibinfo {pages} {155128} (\bibinfo {year} {2018})}\BibitemShut {NoStop}%
\bibitem [{\citenamefont {Romero}\ \emph {et~al.}(2013)\citenamefont {Romero}, \citenamefont {Bingham}, \citenamefont {Forder},\ and\ \citenamefont {Hayward}}]{Fabio_2013}%
  \BibitemOpen
  \bibfield  {author} {\bibinfo {author} {\bibfnamefont {F.~D.}\ \bibnamefont {Romero}}, \bibinfo {author} {\bibfnamefont {P.~A.}\ \bibnamefont {Bingham}}, \bibinfo {author} {\bibfnamefont {S.~D.}\ \bibnamefont {Forder}},\ and\ \bibinfo {author} {\bibfnamefont {M.~A.}\ \bibnamefont {Hayward}},\ }\bibfield  {title} {\bibinfo {title} {Topochemical fluorination of $\mathrm{Sr_{3}(M_{0.5}Ru_{0.5})_{2}O_{7}}$ ($\mathrm{M}$ = $\mathrm{Ti}$, $\mathrm{Mn}$, $\mathrm{Fe}$), $n=2$, ruddlesden--popper~phases},\ }\href {https://doi.org/10.1021/ic400125x} {\bibfield  {journal} {\bibinfo  {journal} {Inorganic Chemistry}\ }\textbf {\bibinfo {volume} {52}},\ \bibinfo {pages} {3388} (\bibinfo {year} {2013})},\ \bibinfo {note} {pMID: 23441869}\BibitemShut {NoStop}%
\bibitem [{\citenamefont {Kageyama}\ \emph {et~al.}(2018)\citenamefont {Kageyama}, \citenamefont {Hayashi}, \citenamefont {Maeda}, \citenamefont {Attfield}, \citenamefont {Hiroi}, \citenamefont {Rondinelli},\ and\ \citenamefont {Poeppelmeier}}]{Kageyama2018}%
  \BibitemOpen
  \bibfield  {author} {\bibinfo {author} {\bibfnamefont {H.}~\bibnamefont {Kageyama}}, \bibinfo {author} {\bibfnamefont {K.}~\bibnamefont {Hayashi}}, \bibinfo {author} {\bibfnamefont {K.}~\bibnamefont {Maeda}}, \bibinfo {author} {\bibfnamefont {J.~P.}\ \bibnamefont {Attfield}}, \bibinfo {author} {\bibfnamefont {Z.}~\bibnamefont {Hiroi}}, \bibinfo {author} {\bibfnamefont {J.~M.}\ \bibnamefont {Rondinelli}},\ and\ \bibinfo {author} {\bibfnamefont {K.~R.}\ \bibnamefont {Poeppelmeier}},\ }\bibfield  {title} {\bibinfo {title} {Expanding frontiers in materials chemistry and physics with multiple anions},\ }\bibfield  {journal} {\bibinfo  {journal} {Nature Communications}\ }\textbf {\bibinfo {volume} {9}},\ \href {https://doi.org/10.1038/s41467-018-02838-4} {10.1038/s41467-018-02838-4} (\bibinfo {year} {2018})\BibitemShut {NoStop}%
\bibitem [{\citenamefont {Acrivos}\ \emph {et~al.}(1994)\citenamefont {Acrivos}, \citenamefont {Lei}, \citenamefont {Jiang}, \citenamefont {Nguyen}, \citenamefont {Metcalf},\ and\ \citenamefont {Honig}}]{Acrivos1994}%
  \BibitemOpen
  \bibfield  {author} {\bibinfo {author} {\bibfnamefont {J.}~\bibnamefont {Acrivos}}, \bibinfo {author} {\bibfnamefont {M.}~\bibnamefont {Lei}}, \bibinfo {author} {\bibfnamefont {C.}~\bibnamefont {Jiang}}, \bibinfo {author} {\bibfnamefont {H.}~\bibnamefont {Nguyen}}, \bibinfo {author} {\bibfnamefont {P.}~\bibnamefont {Metcalf}},\ and\ \bibinfo {author} {\bibfnamefont {J.}~\bibnamefont {Honig}},\ }\bibfield  {title} {\bibinfo {title} {Paramagnetism, antiferromagnetism, and superconductivity in $\mathrm{La}_{2}\mathrm{Ni}\mathrm{O}_{4}$},\ }\href {https://doi.org/https://doi.org/10.1006/jssc.1994.1237} {\bibfield  {journal} {\bibinfo  {journal} {Journal of Solid State Chemistry}\ }\textbf {\bibinfo {volume} {111}},\ \bibinfo {pages} {343} (\bibinfo {year} {1994})}\BibitemShut {NoStop}%
\bibitem [{\citenamefont {Gopalan}\ \emph {et~al.}(1992)\citenamefont {Gopalan}, \citenamefont {McElfresh}, \citenamefont {Ka\ifmmode~\mbox{\c{}}\else \c{}\fi{}kol}, \citenamefont {Spal/ek},\ and\ \citenamefont {Honig}}]{Gopalan.45.249}%
  \BibitemOpen
  \bibfield  {author} {\bibinfo {author} {\bibfnamefont {P.}~\bibnamefont {Gopalan}}, \bibinfo {author} {\bibfnamefont {M.~W.}\ \bibnamefont {McElfresh}}, \bibinfo {author} {\bibfnamefont {Z.}~\bibnamefont {Ka\ifmmode~\mbox{\c{}}\else \c{}\fi{}kol}}, \bibinfo {author} {\bibfnamefont {J.}~\bibnamefont {Spal/ek}},\ and\ \bibinfo {author} {\bibfnamefont {J.~M.}\ \bibnamefont {Honig}},\ }\bibfield  {title} {\bibinfo {title} {Influence of oxygen stoichiometry on the antiferromagnetic ordering of single crystals of $\mathrm{La}_{2}\mathrm{Ni}\mathrm{O}_{4}$},\ }\href {https://doi.org/10.1103/PhysRevB.45.249} {\bibfield  {journal} {\bibinfo  {journal} {Phys. Rev. B}\ }\textbf {\bibinfo {volume} {45}},\ \bibinfo {pages} {249} (\bibinfo {year} {1992})}\BibitemShut {NoStop}%
\bibitem [{\citenamefont {Paulus}\ \emph {et~al.}(2002)\citenamefont {Paulus}, \citenamefont {Cousson}, \citenamefont {Dhalenne}, \citenamefont {Berthon}, \citenamefont {Revcolevschi}, \citenamefont {Hosoya}, \citenamefont {Treutmann}, \citenamefont {Heger},\ and\ \citenamefont {Toquin}}]{PAULUS2002565}%
  \BibitemOpen
  \bibfield  {author} {\bibinfo {author} {\bibfnamefont {W.}~\bibnamefont {Paulus}}, \bibinfo {author} {\bibfnamefont {A.}~\bibnamefont {Cousson}}, \bibinfo {author} {\bibfnamefont {G.}~\bibnamefont {Dhalenne}}, \bibinfo {author} {\bibfnamefont {J.}~\bibnamefont {Berthon}}, \bibinfo {author} {\bibfnamefont {A.}~\bibnamefont {Revcolevschi}}, \bibinfo {author} {\bibfnamefont {S.}~\bibnamefont {Hosoya}}, \bibinfo {author} {\bibfnamefont {W.}~\bibnamefont {Treutmann}}, \bibinfo {author} {\bibfnamefont {G.}~\bibnamefont {Heger}},\ and\ \bibinfo {author} {\bibfnamefont {R.~L.}\ \bibnamefont {Toquin}},\ }\bibfield  {title} {\bibinfo {title} {Neutron diffraction studies of stoichiometric and oxygen intercalated $\mathrm{La}_{2}\mathrm{Ni}\mathrm{O}_{4}$ single crystals},\ }\href {https://doi.org/https://doi.org/10.1016/S1293-2558(02)01299-2} {\bibfield  {journal} {\bibinfo  {journal} {Solid State Sciences}\ }\textbf {\bibinfo {volume} {4}},\ \bibinfo {pages} {565} (\bibinfo {year} {2002})}\BibitemShut {NoStop}%
\bibitem [{\citenamefont {Lander}\ \emph {et~al.}(1989)\citenamefont {Lander}, \citenamefont {Brown}, \citenamefont {Spal/ek},\ and\ \citenamefont {Honig}}]{Lander1989}%
  \BibitemOpen
  \bibfield  {author} {\bibinfo {author} {\bibfnamefont {G.~H.}\ \bibnamefont {Lander}}, \bibinfo {author} {\bibfnamefont {P.~J.}\ \bibnamefont {Brown}}, \bibinfo {author} {\bibfnamefont {J.}~\bibnamefont {Spal/ek}},\ and\ \bibinfo {author} {\bibfnamefont {J.~M.}\ \bibnamefont {Honig}},\ }\bibfield  {title} {\bibinfo {title} {Structural and magnetization density studies of $\mathrm{La}_{2}\mathrm{Ni}\mathrm{O}_{4}$},\ }\href {https://doi.org/10.1103/physrevb.40.4463} {\bibfield  {journal} {\bibinfo  {journal} {Physical Review B}\ }\textbf {\bibinfo {volume} {40}},\ \bibinfo {pages} {4463} (\bibinfo {year} {1989})}\BibitemShut {NoStop}%
\bibitem [{\citenamefont {Wissel}\ \emph {et~al.}(2020{\natexlab{b}})\citenamefont {Wissel}, \citenamefont {Malik}, \citenamefont {Vasala}, \citenamefont {Plana-Ruiz}, \citenamefont {Kolb}, \citenamefont {Slater}, \citenamefont {da~Silva}, \citenamefont {Alff},\ and\ \citenamefont {Clemens}}]{Wissel2020_ChemMat_3160}%
  \BibitemOpen
  \bibfield  {author} {\bibinfo {author} {\bibfnamefont {K.}~\bibnamefont {Wissel}}, \bibinfo {author} {\bibfnamefont {A.~M.}\ \bibnamefont {Malik}}, \bibinfo {author} {\bibfnamefont {S.}~\bibnamefont {Vasala}}, \bibinfo {author} {\bibfnamefont {S.}~\bibnamefont {Plana-Ruiz}}, \bibinfo {author} {\bibfnamefont {U.}~\bibnamefont {Kolb}}, \bibinfo {author} {\bibfnamefont {P.~R.}\ \bibnamefont {Slater}}, \bibinfo {author} {\bibfnamefont {I.}~\bibnamefont {da~Silva}}, \bibinfo {author} {\bibfnamefont {L.}~\bibnamefont {Alff}},\ and\ \bibinfo {author} {\bibfnamefont {O.}~\bibnamefont {Clemens}},\ }\bibfield  {title} {\bibinfo {title} {Topochemical reduction of $\mathrm{La}_{2}\mathrm{NiO}_{3}\mathrm{F}_{2}$: the first ni-based ruddlesden–popper $n = 1$ t-type structure and the impact of reduction on magnetic ordering},\ }\href {https://doi.org/10.1021/acs.chemmater.0c00193} {\bibfield  {journal} {\bibinfo  {journal} {Chemistry of Materials}\ }\textbf {\bibinfo {volume} {32}},\ \bibinfo {pages} {3160} (\bibinfo
  {year} {2020}{\natexlab{b}})}\BibitemShut {NoStop}%
\bibitem [{\citenamefont {Bernardini}\ \emph {et~al.}(2021)\citenamefont {Bernardini}, \citenamefont {Demourgues},\ and\ \citenamefont {Cano}}]{Bernardini2021}%
  \BibitemOpen
  \bibfield  {author} {\bibinfo {author} {\bibfnamefont {F.}~\bibnamefont {Bernardini}}, \bibinfo {author} {\bibfnamefont {A.}~\bibnamefont {Demourgues}},\ and\ \bibinfo {author} {\bibfnamefont {A.}~\bibnamefont {Cano}},\ }\bibfield  {title} {\bibinfo {title} {Single-layer $\mathrm{T}’$-type nickelates: $\mathrm{Ni}^{1+}$ is $\mathrm{Ni}^{1+}$},\ }\href {https://doi.org/10.1103/physrevmaterials.5.l061801} {\bibfield  {journal} {\bibinfo  {journal} {Physical Review Materials}\ }\textbf {\bibinfo {volume} {5}},\ \bibinfo {pages} {l061801} (\bibinfo {year} {2021})}\BibitemShut {NoStop}%
\bibitem [{\citenamefont {Harada}\ \emph {et~al.}(2024)\citenamefont {Harada}, \citenamefont {Charles}, \citenamefont {Koocher}, \citenamefont {Wang}, \citenamefont {Kamp}, \citenamefont {Baxter}, \citenamefont {Poeppelmeier}, \citenamefont {Puggioni},\ and\ \citenamefont {Rondinelli}}]{Harada2024}%
  \BibitemOpen
  \bibfield  {author} {\bibinfo {author} {\bibfnamefont {J.~K.}\ \bibnamefont {Harada}}, \bibinfo {author} {\bibfnamefont {N.}~\bibnamefont {Charles}}, \bibinfo {author} {\bibfnamefont {N.~Z.}\ \bibnamefont {Koocher}}, \bibinfo {author} {\bibfnamefont {Y.}~\bibnamefont {Wang}}, \bibinfo {author} {\bibfnamefont {K.~R.}\ \bibnamefont {Kamp}}, \bibinfo {author} {\bibfnamefont {M.~R.}\ \bibnamefont {Baxter}}, \bibinfo {author} {\bibfnamefont {K.~R.}\ \bibnamefont {Poeppelmeier}}, \bibinfo {author} {\bibfnamefont {D.}~\bibnamefont {Puggioni}},\ and\ \bibinfo {author} {\bibfnamefont {J.~M.}\ \bibnamefont {Rondinelli}},\ }\bibfield  {title} {\bibinfo {title} {Heteroanionic stabilization of $\mathrm{Ni}^{1+}$ with nonplanar coordination in layered nickelates},\ }\href {https://doi.org/10.1103/physrevmaterials.8.024803} {\bibfield  {journal} {\bibinfo  {journal} {Physical Review Materials}\ }\textbf {\bibinfo {volume} {8}},\ \bibinfo {pages} {024803} (\bibinfo {year} {2024})}\BibitemShut {NoStop}%
\bibitem [{\citenamefont {Li}\ \emph {et~al.}(2019)\citenamefont {Li}, \citenamefont {Lee}, \citenamefont {Wang}, \citenamefont {Osada}, \citenamefont {Crossley}, \citenamefont {Lee}, \citenamefont {Cui}, \citenamefont {Hikita},\ and\ \citenamefont {Hwang}}]{Li2019}%
  \BibitemOpen
  \bibfield  {author} {\bibinfo {author} {\bibfnamefont {D.}~\bibnamefont {Li}}, \bibinfo {author} {\bibfnamefont {K.}~\bibnamefont {Lee}}, \bibinfo {author} {\bibfnamefont {B.~Y.}\ \bibnamefont {Wang}}, \bibinfo {author} {\bibfnamefont {M.}~\bibnamefont {Osada}}, \bibinfo {author} {\bibfnamefont {S.}~\bibnamefont {Crossley}}, \bibinfo {author} {\bibfnamefont {H.~R.}\ \bibnamefont {Lee}}, \bibinfo {author} {\bibfnamefont {Y.}~\bibnamefont {Cui}}, \bibinfo {author} {\bibfnamefont {Y.}~\bibnamefont {Hikita}},\ and\ \bibinfo {author} {\bibfnamefont {H.~Y.}\ \bibnamefont {Hwang}},\ }\bibfield  {title} {\bibinfo {title} {Superconductivity in an infinite-layer nickelate},\ }\href {https://doi.org/10.1038/s41586-019-1496-5} {\bibfield  {journal} {\bibinfo  {journal} {Nature}\ }\textbf {\bibinfo {volume} {572}},\ \bibinfo {pages} {624} (\bibinfo {year} {2019})}\BibitemShut {NoStop}%
\bibitem [{\citenamefont {Puphal}\ \emph {et~al.}(2021)\citenamefont {Puphal}, \citenamefont {Wu}, \citenamefont {Fürsich}, \citenamefont {Lee}, \citenamefont {Pakdaman}, \citenamefont {Bruin}, \citenamefont {Nuss}, \citenamefont {Suyolcu}, \citenamefont {van Aken}, \citenamefont {Keimer}, \citenamefont {Isobe},\ and\ \citenamefont {Hepting}}]{Puphal2021}%
  \BibitemOpen
  \bibfield  {author} {\bibinfo {author} {\bibfnamefont {P.}~\bibnamefont {Puphal}}, \bibinfo {author} {\bibfnamefont {Y.-M.}\ \bibnamefont {Wu}}, \bibinfo {author} {\bibfnamefont {K.}~\bibnamefont {Fürsich}}, \bibinfo {author} {\bibfnamefont {H.}~\bibnamefont {Lee}}, \bibinfo {author} {\bibfnamefont {M.}~\bibnamefont {Pakdaman}}, \bibinfo {author} {\bibfnamefont {J.~A.~N.}\ \bibnamefont {Bruin}}, \bibinfo {author} {\bibfnamefont {J.}~\bibnamefont {Nuss}}, \bibinfo {author} {\bibfnamefont {Y.~E.}\ \bibnamefont {Suyolcu}}, \bibinfo {author} {\bibfnamefont {P.~A.}\ \bibnamefont {van Aken}}, \bibinfo {author} {\bibfnamefont {B.}~\bibnamefont {Keimer}}, \bibinfo {author} {\bibfnamefont {M.}~\bibnamefont {Isobe}},\ and\ \bibinfo {author} {\bibfnamefont {M.}~\bibnamefont {Hepting}},\ }\bibfield  {title} {\bibinfo {title} {Topotactic transformation of single crystals: From perovskite to infinite-layer nickelates},\ }\bibfield  {journal} {\bibinfo  {journal} {Science Advances}\ }\textbf {\bibinfo {volume} {7}},\
  \href {https://doi.org/10.1126/sciadv.abl8091} {10.1126/sciadv.abl8091} (\bibinfo {year} {2021})\BibitemShut {NoStop}%
\bibitem [{\citenamefont {Wu}\ \emph {et~al.}(2023)\citenamefont {Wu}, \citenamefont {Puphal}, \citenamefont {Lee}, \citenamefont {Nuss}, \citenamefont {Isobe}, \citenamefont {Keimer}, \citenamefont {Hepting}, \citenamefont {Suyolcu},\ and\ \citenamefont {van Aken}}]{Wu2023}%
  \BibitemOpen
  \bibfield  {author} {\bibinfo {author} {\bibfnamefont {Y.-M.}\ \bibnamefont {Wu}}, \bibinfo {author} {\bibfnamefont {P.}~\bibnamefont {Puphal}}, \bibinfo {author} {\bibfnamefont {H.}~\bibnamefont {Lee}}, \bibinfo {author} {\bibfnamefont {J.}~\bibnamefont {Nuss}}, \bibinfo {author} {\bibfnamefont {M.}~\bibnamefont {Isobe}}, \bibinfo {author} {\bibfnamefont {B.}~\bibnamefont {Keimer}}, \bibinfo {author} {\bibfnamefont {M.}~\bibnamefont {Hepting}}, \bibinfo {author} {\bibfnamefont {Y.~E.}\ \bibnamefont {Suyolcu}},\ and\ \bibinfo {author} {\bibfnamefont {P.~A.}\ \bibnamefont {van Aken}},\ }\bibfield  {title} {\bibinfo {title} {{Topotactically induced oxygen vacancy order in nickelate single crystals}},\ }\href {https://doi.org/10.1103/PhysRevMaterials.7.053609} {\bibfield  {journal} {\bibinfo  {journal} {Phys. Rev. Mater.}\ }\textbf {\bibinfo {volume} {7}},\ \bibinfo {pages} {053609} (\bibinfo {year} {2023})}\BibitemShut {NoStop}%
\bibitem [{\citenamefont {Suyolcu}\ \emph {et~al.}(2025)\citenamefont {Suyolcu}, \citenamefont {Puphal},\ and\ \citenamefont {Hepting}}]{Suyolcu2025}%
  \BibitemOpen
  \bibfield  {author} {\bibinfo {author} {\bibfnamefont {Y.~E.}\ \bibnamefont {Suyolcu}}, \bibinfo {author} {\bibfnamefont {P.}~\bibnamefont {Puphal}},\ and\ \bibinfo {author} {\bibfnamefont {M.}~\bibnamefont {Hepting}},\ }\bibfield  {title} {\bibinfo {title} {Three generations of infinite-layer nickelate crystals},\ }\bibfield  {journal} {\bibinfo  {journal} {MRS Communications}\ }\href {https://doi.org/10.1557/s43579-025-00689-x} {10.1557/s43579-025-00689-x} (\bibinfo {year} {2025})\BibitemShut {NoStop}%
\bibitem [{\citenamefont {Wu}\ \emph {et~al.}(2024)\citenamefont {Wu}, \citenamefont {Puphal}, \citenamefont {Isobe}, \citenamefont {Keimer}, \citenamefont {Hepting}, \citenamefont {Suyolcu},\ and\ \citenamefont {van Aken}}]{Wu2024}%
  \BibitemOpen
  \bibfield  {author} {\bibinfo {author} {\bibfnamefont {Y.-M.}\ \bibnamefont {Wu}}, \bibinfo {author} {\bibfnamefont {P.}~\bibnamefont {Puphal}}, \bibinfo {author} {\bibfnamefont {M.}~\bibnamefont {Isobe}}, \bibinfo {author} {\bibfnamefont {B.}~\bibnamefont {Keimer}}, \bibinfo {author} {\bibfnamefont {M.}~\bibnamefont {Hepting}}, \bibinfo {author} {\bibfnamefont {Y.~E.}\ \bibnamefont {Suyolcu}},\ and\ \bibinfo {author} {\bibfnamefont {P.~A.}\ \bibnamefont {van Aken}},\ }\bibfield  {title} {\bibinfo {title} {{Unraveling nano-scale effects of topotactic reduction in LaNiO$_2$ crystals}},\ }\bibfield  {journal} {\bibinfo  {journal} {APL Materials}\ }\textbf {\bibinfo {volume} {12}},\ \href {https://doi.org/10.1063/5.0227732} {10.1063/5.0227732} (\bibinfo {year} {2024})\BibitemShut {NoStop}%
\bibitem [{\citenamefont {Tamura}\ \emph {et~al.}(1993)\citenamefont {Tamura}, \citenamefont {Hayashi},\ and\ \citenamefont {Ueda}}]{Tamura1993}%
  \BibitemOpen
  \bibfield  {author} {\bibinfo {author} {\bibfnamefont {H.}~\bibnamefont {Tamura}}, \bibinfo {author} {\bibfnamefont {A.}~\bibnamefont {Hayashi}},\ and\ \bibinfo {author} {\bibfnamefont {Y.}~\bibnamefont {Ueda}},\ }\bibfield  {title} {\bibinfo {title} {Phase diagram of $\mathrm{La}_{2}\mathrm{Ni}\mathrm{O}_{4+\delta}$ ($0 \leq \delta \leq 0.18$)},\ }\href {https://doi.org/10.1016/0921-4534(93)90636-5} {\bibfield  {journal} {\bibinfo  {journal} {Physica C: Superconductivity}\ }\textbf {\bibinfo {volume} {216}},\ \bibinfo {pages} {83} (\bibinfo {year} {1993})}\BibitemShut {NoStop}%
\bibitem [{\citenamefont {Nowroozi}\ \emph {et~al.}(2020)\citenamefont {Nowroozi}, \citenamefont {Wissel}, \citenamefont {Donzelli}, \citenamefont {Hosseinpourkahvaz}, \citenamefont {Plana-Ruiz}, \citenamefont {Kolb}, \citenamefont {Schoch}, \citenamefont {Bauer}, \citenamefont {Malik}, \citenamefont {Rohrer}, \citenamefont {Ivlev}, \citenamefont {Kraus},\ and\ \citenamefont {Clemens}}]{Nowroozi2020_CommMat_27}%
  \BibitemOpen
  \bibfield  {author} {\bibinfo {author} {\bibfnamefont {M.~A.}\ \bibnamefont {Nowroozi}}, \bibinfo {author} {\bibfnamefont {K.}~\bibnamefont {Wissel}}, \bibinfo {author} {\bibfnamefont {M.}~\bibnamefont {Donzelli}}, \bibinfo {author} {\bibfnamefont {N.}~\bibnamefont {Hosseinpourkahvaz}}, \bibinfo {author} {\bibfnamefont {S.}~\bibnamefont {Plana-Ruiz}}, \bibinfo {author} {\bibfnamefont {U.}~\bibnamefont {Kolb}}, \bibinfo {author} {\bibfnamefont {R.}~\bibnamefont {Schoch}}, \bibinfo {author} {\bibfnamefont {M.}~\bibnamefont {Bauer}}, \bibinfo {author} {\bibfnamefont {A.~M.}\ \bibnamefont {Malik}}, \bibinfo {author} {\bibfnamefont {J.}~\bibnamefont {Rohrer}}, \bibinfo {author} {\bibfnamefont {S.}~\bibnamefont {Ivlev}}, \bibinfo {author} {\bibfnamefont {F.}~\bibnamefont {Kraus}},\ and\ \bibinfo {author} {\bibfnamefont {O.}~\bibnamefont {Clemens}},\ }\bibfield  {title} {\bibinfo {title} {High cycle life all-solid-state fluoride ion battery with $\mathrm{La}_{2}\mathrm{Ni}\mathrm{O}_{4+\delta}$ high voltage
  cathode},\ }\bibfield  {journal} {\bibinfo  {journal} {Communications Materials}\ }\textbf {\bibinfo {volume} {1}},\ \href {https://doi.org/10.1038/s43246-020-0030-5} {10.1038/s43246-020-0030-5} (\bibinfo {year} {2020})\BibitemShut {NoStop}%
\bibitem [{\citenamefont {Rodríguez-Carvajal}\ \emph {et~al.}(1991)\citenamefont {Rodríguez-Carvajal}, \citenamefont {Fernández-Díaz},\ and\ \citenamefont {Martínez}}]{RodriguezCarvajal1991_JPCM_3215}%
  \BibitemOpen
  \bibfield  {author} {\bibinfo {author} {\bibfnamefont {J.}~\bibnamefont {Rodríguez-Carvajal}}, \bibinfo {author} {\bibfnamefont {M.~T.}\ \bibnamefont {Fernández-Díaz}},\ and\ \bibinfo {author} {\bibfnamefont {J.~L.}\ \bibnamefont {Martínez}},\ }\bibfield  {title} {\bibinfo {title} {Neutron diffraction study on structural and magnetic properties of $\mathrm{La}_{2}\mathrm{Ni}\mathrm{O}_{4}$},\ }\href {https://doi.org/10.1088/0953-8984/3/19/002} {\bibfield  {journal} {\bibinfo  {journal} {Journal of Physics: Condensed Matter}\ }\textbf {\bibinfo {volume} {3}},\ \bibinfo {pages} {3215} (\bibinfo {year} {1991})}\BibitemShut {NoStop}%
\bibitem [{\citenamefont {Jacobs}\ \emph {et~al.}(2025)\citenamefont {Jacobs}, \citenamefont {Bivour}, \citenamefont {Sikolenko}, \citenamefont {Kohlmann}, \citenamefont {Hansen}, \citenamefont {Hester}, \citenamefont {Xu}, \citenamefont {Schmedt auf~der Günne},\ and\ \citenamefont {Ebbinghaus}}]{Jacobs2025}%
  \BibitemOpen
  \bibfield  {author} {\bibinfo {author} {\bibfnamefont {J.}~\bibnamefont {Jacobs}}, \bibinfo {author} {\bibfnamefont {A.}~\bibnamefont {Bivour}}, \bibinfo {author} {\bibfnamefont {V.}~\bibnamefont {Sikolenko}}, \bibinfo {author} {\bibfnamefont {H.}~\bibnamefont {Kohlmann}}, \bibinfo {author} {\bibfnamefont {T.~C.}\ \bibnamefont {Hansen}}, \bibinfo {author} {\bibfnamefont {J.~R.}\ \bibnamefont {Hester}}, \bibinfo {author} {\bibfnamefont {K.}~\bibnamefont {Xu}}, \bibinfo {author} {\bibfnamefont {J.}~\bibnamefont {Schmedt auf~der Günne}},\ and\ \bibinfo {author} {\bibfnamefont {S.~G.}\ \bibnamefont {Ebbinghaus}},\ }\bibfield  {title} {\bibinfo {title} {Unveiling the fluorination pathway of ruddlesden–popper oxyfluorides: A comprehensive in situ x-ray and neutron diffraction study},\ }\href {https://doi.org/10.1021/jacs.4c18187} {\bibfield  {journal} {\bibinfo  {journal} {Journal of the American Chemical Society}\ }\textbf {\bibinfo {volume} {147}},\ \bibinfo {pages} {9739} (\bibinfo {year}
  {2025})}\BibitemShut {NoStop}%
\bibitem [{\citenamefont {Jacobs}\ \emph {et~al.}(2024)\citenamefont {Jacobs}, \citenamefont {Wang}, \citenamefont {Marques},\ and\ \citenamefont {Ebbinghaus}}]{Jacobs2024}%
  \BibitemOpen
  \bibfield  {author} {\bibinfo {author} {\bibfnamefont {J.}~\bibnamefont {Jacobs}}, \bibinfo {author} {\bibfnamefont {H.-C.}\ \bibnamefont {Wang}}, \bibinfo {author} {\bibfnamefont {M.~A.~L.}\ \bibnamefont {Marques}},\ and\ \bibinfo {author} {\bibfnamefont {S.~G.}\ \bibnamefont {Ebbinghaus}},\ }\bibfield  {title} {\bibinfo {title} {Ruddlesden–popper oxyfluorides $\mathrm{La}_{2}\mathrm{Ni}_{1-x}\mathrm{Cu}_{x}\mathrm{O}_{3}\mathrm{F}_{2}$ ($0 \leq x \leq 1$): Impact of the $\mathrm{Ni/Cu}$ ratio on the thermal stability and magnetic properties},\ }\href {https://doi.org/10.1021/acs.inorgchem.4c01330} {\bibfield  {journal} {\bibinfo  {journal} {Inorganic Chemistry}\ }\textbf {\bibinfo {volume} {63}},\ \bibinfo {pages} {11317} (\bibinfo {year} {2024})}\BibitemShut {NoStop}%
\bibitem [{\citenamefont {Puphal}\ \emph {et~al.}(2023{\natexlab{b}})\citenamefont {Puphal}, \citenamefont {Akbar}, \citenamefont {Hepting}, \citenamefont {Goering}, \citenamefont {Isobe}, \citenamefont {Nugroho},\ and\ \citenamefont {Keimer}}]{Puphal2023}%
  \BibitemOpen
  \bibfield  {author} {\bibinfo {author} {\bibfnamefont {P.}~\bibnamefont {Puphal}}, \bibinfo {author} {\bibfnamefont {M.~Y.~P.}\ \bibnamefont {Akbar}}, \bibinfo {author} {\bibfnamefont {M.}~\bibnamefont {Hepting}}, \bibinfo {author} {\bibfnamefont {E.}~\bibnamefont {Goering}}, \bibinfo {author} {\bibfnamefont {M.}~\bibnamefont {Isobe}}, \bibinfo {author} {\bibfnamefont {A.~A.}\ \bibnamefont {Nugroho}},\ and\ \bibinfo {author} {\bibfnamefont {B.}~\bibnamefont {Keimer}},\ }\bibfield  {title} {\bibinfo {title} {Single crystal synthesis, structure, and magnetism of $\mathrm{Pb}_{10-x}\mathrm{Cu}_x(\mathrm{PO}_4)_6\mathrm{O}$},\ }\bibfield  {journal} {\bibinfo  {journal} {APL Materials}\ }\textbf {\bibinfo {volume} {11}},\ \href {https://doi.org/10.1063/5.0172755} {10.1063/5.0172755} (\bibinfo {year} {2023}{\natexlab{b}})\BibitemShut {NoStop}%
\bibitem [{\citenamefont {Krieger}\ \emph {et~al.}(2025)\citenamefont {Krieger}, \citenamefont {Hicken}, \citenamefont {Luetkens}, \citenamefont {Kremer},\ and\ \citenamefont {Puphal}}]{Krieger2025}%
  \BibitemOpen
  \bibfield  {author} {\bibinfo {author} {\bibfnamefont {J.~A.}\ \bibnamefont {Krieger}}, \bibinfo {author} {\bibfnamefont {T.~J.}\ \bibnamefont {Hicken}}, \bibinfo {author} {\bibfnamefont {H.}~\bibnamefont {Luetkens}}, \bibinfo {author} {\bibfnamefont {R.~K.}\ \bibnamefont {Kremer}},\ and\ \bibinfo {author} {\bibfnamefont {P.}~\bibnamefont {Puphal}},\ }\bibfield  {title} {\bibinfo {title} {Exerting chemical pressure on the kagome lattice as frustration control in the kapellasite family ${A}$$\mathrm{Cu}_3(\mathrm{OH})_{6+x}(\mathrm{Cl,Br})_{3-x}$},\ }\bibfield  {journal} {\bibinfo  {journal} {Physical Review Research}\ }\textbf {\bibinfo {volume} {7}},\ \href {https://doi.org/10.1103/mgyp-pcpl} {10.1103/mgyp-pcpl} (\bibinfo {year} {2025})\BibitemShut {NoStop}%
\bibitem [{\citenamefont {Hayashida}\ \emph {et~al.}(2024)\citenamefont {Hayashida}, \citenamefont {Sundaramurthy}, \citenamefont {Puphal}, \citenamefont {Garcia-Fernandez}, \citenamefont {Zhou}, \citenamefont {Fenk}, \citenamefont {Isobe}, \citenamefont {Minola}, \citenamefont {Wu}, \citenamefont {Suyolcu}, \citenamefont {van Aken}, \citenamefont {Keimer},\ and\ \citenamefont {Hepting}}]{Hayashida2024}%
  \BibitemOpen
  \bibfield  {author} {\bibinfo {author} {\bibfnamefont {S.}~\bibnamefont {Hayashida}}, \bibinfo {author} {\bibfnamefont {V.}~\bibnamefont {Sundaramurthy}}, \bibinfo {author} {\bibfnamefont {P.}~\bibnamefont {Puphal}}, \bibinfo {author} {\bibfnamefont {M.}~\bibnamefont {Garcia-Fernandez}}, \bibinfo {author} {\bibfnamefont {K.-J.}\ \bibnamefont {Zhou}}, \bibinfo {author} {\bibfnamefont {B.}~\bibnamefont {Fenk}}, \bibinfo {author} {\bibfnamefont {M.}~\bibnamefont {Isobe}}, \bibinfo {author} {\bibfnamefont {M.}~\bibnamefont {Minola}}, \bibinfo {author} {\bibfnamefont {Y.-M.}\ \bibnamefont {Wu}}, \bibinfo {author} {\bibfnamefont {Y.~E.}\ \bibnamefont {Suyolcu}}, \bibinfo {author} {\bibfnamefont {P.~A.}\ \bibnamefont {van Aken}}, \bibinfo {author} {\bibfnamefont {B.}~\bibnamefont {Keimer}},\ and\ \bibinfo {author} {\bibfnamefont {M.}~\bibnamefont {Hepting}},\ }\bibfield  {title} {\bibinfo {title} {{Investigation of spin excitations and charge order in bulk crystals of the infinite-layer nickelate
  ${\mathrm{LaNiO}}_{2}$}},\ }\href {https://doi.org/10.1103/PhysRevB.109.235106} {\bibfield  {journal} {\bibinfo  {journal} {Phys. Rev. B}\ }\textbf {\bibinfo {volume} {109}},\ \bibinfo {pages} {235106} (\bibinfo {year} {2024})}\BibitemShut {NoStop}%
\bibitem [{\citenamefont {Scimeca}\ \emph {et~al.}(2018)\citenamefont {Scimeca}, \citenamefont {Bischetti}, \citenamefont {Lamsira}, \citenamefont {Bonfiglio},\ and\ \citenamefont {Bonanno}}]{Scimeca2018_EJH_2841}%
  \BibitemOpen
  \bibfield  {author} {\bibinfo {author} {\bibfnamefont {M.}~\bibnamefont {Scimeca}}, \bibinfo {author} {\bibfnamefont {S.}~\bibnamefont {Bischetti}}, \bibinfo {author} {\bibfnamefont {H.~K.}\ \bibnamefont {Lamsira}}, \bibinfo {author} {\bibfnamefont {R.}~\bibnamefont {Bonfiglio}},\ and\ \bibinfo {author} {\bibfnamefont {E.}~\bibnamefont {Bonanno}},\ }\bibfield  {title} {\bibinfo {title} {Energy dispersive x-ray (edx) microanalysis: A powerful tool in biomedical research and diagnosis},\ }\href {https://doi.org/10.4081/ejh.2018.2841} {\bibfield  {journal} {\bibinfo  {journal} {European Journal of Histochemistry}\ }\textbf {\bibinfo {volume} {62}},\ \bibinfo {pages} {2841} (\bibinfo {year} {2018})}\BibitemShut {NoStop}%
\bibitem [{\citenamefont {Arumugam}\ \emph {et~al.}(2019)\citenamefont {Arumugam}, \citenamefont {Sivaprakash}, \citenamefont {Dixit}, \citenamefont {Chaurasiya}, \citenamefont {Govindaraj}, \citenamefont {Sathiskumar}, \citenamefont {Chatterjee},\ and\ \citenamefont {Suryanarayanan}}]{Arumugam2019}%
  \BibitemOpen
  \bibfield  {author} {\bibinfo {author} {\bibfnamefont {S.}~\bibnamefont {Arumugam}}, \bibinfo {author} {\bibfnamefont {P.}~\bibnamefont {Sivaprakash}}, \bibinfo {author} {\bibfnamefont {A.}~\bibnamefont {Dixit}}, \bibinfo {author} {\bibfnamefont {R.}~\bibnamefont {Chaurasiya}}, \bibinfo {author} {\bibfnamefont {L.}~\bibnamefont {Govindaraj}}, \bibinfo {author} {\bibfnamefont {M.}~\bibnamefont {Sathiskumar}}, \bibinfo {author} {\bibfnamefont {S.}~\bibnamefont {Chatterjee}},\ and\ \bibinfo {author} {\bibfnamefont {R.}~\bibnamefont {Suryanarayanan}},\ }\bibfield  {title} {\bibinfo {title} {{Complex magnetic structure and magnetocapacitance response in a non-oxide NiF$_2$ system}},\ }\bibfield  {journal} {\bibinfo  {journal} {Scientific Reports}\ }\textbf {\bibinfo {volume} {9}},\ \href {https://doi.org/10.1038/s41598-019-39083-8} {10.1038/s41598-019-39083-8} (\bibinfo {year} {2019})\BibitemShut {NoStop}%
\bibitem [{\citenamefont {Ferenc~Segedin}\ \emph {et~al.}(2026)\citenamefont {Ferenc~Segedin}, \citenamefont {Kim}, \citenamefont {LaBollita}, \citenamefont {Taylor}, \citenamefont {Baek}, \citenamefont {Sung}, \citenamefont {Turkiewicz}, \citenamefont {Pan}, \citenamefont {Jiang}, \citenamefont {Bambrick-Santoyo}, \citenamefont {Schwaigert}, \citenamefont {Kim}, \citenamefont {Tenneti}, \citenamefont {Grutter}, \citenamefont {Muramoto}, \citenamefont {N’Diaye}, \citenamefont {El~Baggari}, \citenamefont {Walko}, \citenamefont {Brooks}, \citenamefont {Botana}, \citenamefont {Schlom}, \citenamefont {Zhou},\ and\ \citenamefont {Mundy}}]{FerencSegedin2026}%
  \BibitemOpen
  \bibfield  {author} {\bibinfo {author} {\bibfnamefont {D.}~\bibnamefont {Ferenc~Segedin}}, \bibinfo {author} {\bibfnamefont {J.}~\bibnamefont {Kim}}, \bibinfo {author} {\bibfnamefont {H.}~\bibnamefont {LaBollita}}, \bibinfo {author} {\bibfnamefont {N.~K.}\ \bibnamefont {Taylor}}, \bibinfo {author} {\bibfnamefont {K.-Y.}\ \bibnamefont {Baek}}, \bibinfo {author} {\bibfnamefont {S.~H.}\ \bibnamefont {Sung}}, \bibinfo {author} {\bibfnamefont {A.~B.}\ \bibnamefont {Turkiewicz}}, \bibinfo {author} {\bibfnamefont {G.~A.}\ \bibnamefont {Pan}}, \bibinfo {author} {\bibfnamefont {A.~Y.}\ \bibnamefont {Jiang}}, \bibinfo {author} {\bibfnamefont {M.}~\bibnamefont {Bambrick-Santoyo}}, \bibinfo {author} {\bibfnamefont {T.}~\bibnamefont {Schwaigert}}, \bibinfo {author} {\bibfnamefont {C.~K.}\ \bibnamefont {Kim}}, \bibinfo {author} {\bibfnamefont {A.}~\bibnamefont {Tenneti}}, \bibinfo {author} {\bibfnamefont {A.~J.}\ \bibnamefont {Grutter}}, \bibinfo {author} {\bibfnamefont {S.}~\bibnamefont {Muramoto}}, \bibinfo {author}
  {\bibfnamefont {A.~T.}\ \bibnamefont {N’Diaye}}, \bibinfo {author} {\bibfnamefont {I.}~\bibnamefont {El~Baggari}}, \bibinfo {author} {\bibfnamefont {D.~A.}\ \bibnamefont {Walko}}, \bibinfo {author} {\bibfnamefont {C.~M.}\ \bibnamefont {Brooks}}, \bibinfo {author} {\bibfnamefont {A.~S.}\ \bibnamefont {Botana}}, \bibinfo {author} {\bibfnamefont {D.~G.}\ \bibnamefont {Schlom}}, \bibinfo {author} {\bibfnamefont {H.}~\bibnamefont {Zhou}},\ and\ \bibinfo {author} {\bibfnamefont {J.~A.}\ \bibnamefont {Mundy}},\ }\bibfield  {title} {\bibinfo {title} {Topochemical oxidation of ruddlesden–popper nickelates reveals distinct structural family: Oxygen-intercalated layered perovskites},\ }\href {https://doi.org/10.1021/jacs.5c12712} {\bibfield  {journal} {\bibinfo  {journal} {Journal of the American Chemical Society}\ }\textbf {\bibinfo {volume} {148}},\ \bibinfo {pages} {5873} (\bibinfo {year} {2026})}\BibitemShut {NoStop}%
\bibitem [{\citenamefont {Ko}\ \emph {et~al.}(2024)\citenamefont {Ko}, \citenamefont {Yu}, \citenamefont {Liu}, \citenamefont {Bhatt}, \citenamefont {Li}, \citenamefont {Thampy}, \citenamefont {Kuo}, \citenamefont {Wang}, \citenamefont {Lee}, \citenamefont {Lee}, \citenamefont {Lee}, \citenamefont {Goodge}, \citenamefont {Muller},\ and\ \citenamefont {Hwang}}]{Ko2024}%
  \BibitemOpen
  \bibfield  {author} {\bibinfo {author} {\bibfnamefont {E.~K.}\ \bibnamefont {Ko}}, \bibinfo {author} {\bibfnamefont {Y.}~\bibnamefont {Yu}}, \bibinfo {author} {\bibfnamefont {Y.}~\bibnamefont {Liu}}, \bibinfo {author} {\bibfnamefont {L.}~\bibnamefont {Bhatt}}, \bibinfo {author} {\bibfnamefont {J.}~\bibnamefont {Li}}, \bibinfo {author} {\bibfnamefont {V.}~\bibnamefont {Thampy}}, \bibinfo {author} {\bibfnamefont {C.-T.}\ \bibnamefont {Kuo}}, \bibinfo {author} {\bibfnamefont {B.~Y.}\ \bibnamefont {Wang}}, \bibinfo {author} {\bibfnamefont {Y.}~\bibnamefont {Lee}}, \bibinfo {author} {\bibfnamefont {K.}~\bibnamefont {Lee}}, \bibinfo {author} {\bibfnamefont {J.-S.}\ \bibnamefont {Lee}}, \bibinfo {author} {\bibfnamefont {B.~H.}\ \bibnamefont {Goodge}}, \bibinfo {author} {\bibfnamefont {D.~A.}\ \bibnamefont {Muller}},\ and\ \bibinfo {author} {\bibfnamefont {H.~Y.}\ \bibnamefont {Hwang}},\ }\bibfield  {title} {\bibinfo {title} {Signatures of ambient pressure superconductivity in thin film
  $\mathrm{La}_{3}\mathrm{Ni}_{2}\mathrm{O}_{7}$},\ }\href {https://doi.org/10.1038/s41586-024-08525-3} {\bibfield  {journal} {\bibinfo  {journal} {Nature}\ }\textbf {\bibinfo {volume} {638}},\ \bibinfo {pages} {935} (\bibinfo {year} {2024})}\BibitemShut {NoStop}%
\end{thebibliography}%

\section*{acknowledgments}
We thank the Solid State Spectroscopy department for the use of their SQUID and PPMS systems. H.Y. gratefully acknowledges the financial support of the Study Abroad Postgraduate Education Scholarship (YLSY) awarded by the Republic of Türkiye Ministry of National Education.

\section*{Author Contributions}
H.Y.: Conducted the synthesis of the single crystals, performed and designed fluorination experiments, conducted diffraction measurements, SEM and EDX measurements, as well as all analysis under supervision of P.P. and O.C.\\
O.C.: Supervised the project.\\
P.P.: Supervised the project as well as measured magnetization.\\
HY and P.P. contributed equally to writing and revising the manuscript. All authors read the final version of the manuscript and contributed to it by corrections and/or suggestions.

\section*{Competing Interests}
The authors declare no competing interests.

\end{document}